%% file: Booth1.tex
\documentclass[usenatbib]{mn2e}

\usepackage{graphicx}	
\usepackage{amssymb}

\input{aas_macros.tex}

\begin{document}

\title[The History of the Solar System's Debris Disc]{The History of the Solar System's Debris Disc: Observable Properties of the Kuiper Belt}
\author[M. Booth, M. C. Wyatt, A. Morbidelli, A. Moro-Mart{\'{\i}}n \& H. F. Levison]{Mark Booth$^{1}$\thanks{E-mail: mbooth@ast.cam.ac.uk}, Mark C. Wyatt$^{1}$, Alessandro Morbidelli$^{2}$, Amaya Moro-Mart{\'{\i}}n$^{3,4}$ \newauthor and Harold F. Levison$^{5}$\\
$^{1}$Institute of Astronomy, Madingley Rd, Cambridge CB3 0HA, UK \\
$^{2}$Observatoire de la C\^ote d'Azur, Nice, France \\
$^{3}$Centro de Astrobiologia - CSIC/INTA, 28850 Torrej\'on de Ardoz, Madrid, Spain \\
$^{4}$Department of Astrophysical Sciences, Peyton Hall, Ivy Lane, Princeton University, Princeton, NJ 08544, USA\\
$^{5}$Department of Space Studies, Southwest Research Institute, Boulder, CO 80302, USA
}

\date{Accepted 2009 June 18.  Received 2009 June 16; in original form 2009 April 28}
\pubyear{2009}

\maketitle

\begin{abstract}
The Nice model of \citet{gomes05} suggests that the migration of the giant planets caused a planetesimal clearing event which led to the Late Heavy Bombardment (LHB) at 880~Myr. Here we investigate the IR emission from the Kuiper belt during the history of the Solar System as described by the Nice model. We describe a method for easily converting the results of n-body planetesimal simulations into observational properties (assuming black-body grains and a single size distribution) and further modify this method to improve its realism (using realistic grain properties and a three-phase size distribution). We compare our results with observed debris discs and evaluate the plausibility of detecting an LHB-like process in extrasolar systems. Recent surveys have shown that 4\% of stars exhibit 24~$\mu$m excess and 16\% exhibit 70~$\mu$m excess. We show that the Solar System would have been amongst the brightest of these systems before the LHB at both 24 and 70~$\mu$m. We find a significant increase in 24~$\mu$m emission during the LHB, which rapidly drops off and becomes undetectable within 30~Myr, whereas the 70~$\mu$m emission remains detectable until 360~Myr after the LHB. Comparison with the statistics of debris disc evolution shows that such depletion events must be rare occurring around less than 12\% of Sun-like stars and with this level of incidence we would expect approximately 1 of the 413 Sun-like, field stars so far detected to have a 24~$\mu$m excess to be currently going through an LHB. We also find that collisional processes are important in the Solar System before the LHB and that parameters for weak Kuiper belt objects are inconsistent with the Nice model interpretation of the LHB.

\end{abstract}


\section{Introduction}
\label{s:intro}
Over the past couple of decades, an increasing number of stars have been found to be orbited by discs of planetesimals and dust known as debris discs. As more and more discs are discovered it becomes possible to start building up a picture of how these debris discs evolve over time \citep[see][for a review]{wyatt08}. Recent surveys \citep[e.g.][]{hillenbrand08,trilling08,carpenter09} have shown that the number of Sun-like stars that have been observed with 24~$\mu$m emission (produced by hot dust) decreases with age, but the number of stars with 70~$\mu$m emission (produced by cold dust) remains approximately constant with age. These observations generally agree with models suggesting that debris discs evolve in steady-state becoming collisionally depleted over time \citep{lohne08}, although there are a few exceptions that have much more hot dust than would be expected from these collisional arguments \citep{wyatt07}.

The Solar System has its own debris disc, with the majority of its mass concentrated in the asteroid belt and Kuiper belt. These belts correspond to the hot dust and cold dust seen around other stars, but our own disc is much less massive than these observed discs \citep{moro08}. Simulations of accretion in the Kuiper belt and the formation of binary Kuiper belt objects (KBOs) suggests that the original Kuiper belt must have been much more massive for the largest objects to form \citep[e.g.][]{stern96a,chiang07}, which leads to the `missing mass problem' of the Kuiper belt as this mass deficit cannot be explained by collisional processes alone. 

One model that does explain the missing mass of the Kuiper belt -- along with the orbits of the giant planets and various other details of the structure of the Solar System -- is the Nice model \citep{gomes05,tsiganis05,morbidelli05,levison08}. The Nice model was designed to explain the current orbital elements of the outermost planets \citep{tsiganis05}. It is based on the idea that the gas giants formed much closer together. Due to interactions with the planetesimal disc, Saturn, Neptune and Uranus migrated outwards and Jupiter migrated slightly inwards. When Jupiter and Saturn crossed their 2:1 mean motion resonance (MMR) the system became temporarily destabilised, affecting the orbital elements of the gas giants. As Neptune moved out into the Kuiper belt, it dynamically excited the orbits of many of the KBOs, causing them to evolve on to cometary orbits and impact the terrestrial planets and moons. As the planets' orbits evolved, secular resonance sweeping would have excited the orbits of many of the asteroids \citep{gomes97}, thus also causing a bombardment of asteroids on the planets and moons of the inner Solar System. Hence, the Nice model also explains the Late Heavy Bombardment (LHB) of the Moon -- a period of intense bombardment in which most of the craters on the Moon were formed, which occurred around 3.9 billion years ago \citep{tera74}, of which the latest impactors were most likely main belt asteroids \citep{kring02,strom05}.

This period of intense bombardment and dynamical depletion of the Kuiper belt is likely to have had a significant effect on the observable properties of the debris disc of the Solar System. In this paper we investigate this effect by converting the distributions of planetesimal mass from the Nice model into distributions of emitting surface area to discover how the Solar System would have appeared to a distant observer during its history. Although the Solar System's debris disc has a number of components, for this paper we concentrate on the changes to the Kuiper belt and how this would have affected the observable properties of the Solar System. In section \ref{s:model} we describe the Nice model data and the simple analytical model applied to it, which uses the assumption of black-body grains and a single slope size distribution. In section \ref{s:pp} we look into relaxing the assumptions of black-body emission and a single slope size distribution to discover how a more realistic model changes our initial conclusions. Our final conclusions are given in section \ref{s:con}.

\section{Modelling}
\label{s:model}

\subsection{The Nice model data}
\label{s:ndata}
Our model is based on planetesimal data from one of the Nice model runs \citep{gomes05}. In the Nice model, the simulation begins with effectively 10,000 particles, each representing $1.05\times10^{-8}$~M$_{\odot}$ of Kuiper belt objects. Any particles that reached a heliocentric distance of 1000 AU or evolved onto orbits with perihelion, $q<1$ AU were removed from the simulation. The data covers a 1.2~Gyr time period starting at the time at which the gas disc dissipates. \citet{gomes05} ran 8 simulations with varying disc inner radius. The data used here is for the run with a disc inner edge of $\sim$15.5 AU, which places the LHB at 879~Myr, close to the $\sim$700~Myr given by the analysis of \citet{strom05}. This is also the most realistic of the \citet{gomes05} runs since particles within $\sim$15.3~AU have dynamical lifetimes shorter than the gas disc lifetime showing that they would have disappeared by the time the gas disc dissipates. This run starts with an initial disc mass of 35~M$_{\oplus}$ and has 24~M$_{\oplus}$ at the time of the LHB. If the disc is less massive than this, Jupiter and Saturn do not cross their 2:1 MMR and there is no LHB. If the disc is more massive then the final separation of Jupiter and Saturn is much larger than it is today.

\subsection{Mass evolution}
\label{s:mev}
Using the orbital elements of the Nice model particles we can then calculate the position of the particles at each time-step. For now it is the one-dimensional distribution that we are interested in and so the particles are separated into radius bins to allow us to determine the mass distribution in the system. Due to the small number of particles in the simulation, the evolution of the mass distribution appears stochastic in each 1~Myr time-step. To smooth out the evolution and make it more realistic, two changes are made. Firstly, each particle is replaced by 10 particles, each of which is one tenth of the original mass, and these particles are spread uniformly around the orbit in mean anomaly to simulate the entire range of radii that particle would have passed through during this time period. By doing this we lose any resonant structures present in the data but increase the resolution of the model. Secondly, the mass distribution at each time-step is averaged over 5 time-steps ($\sim$5~Myr). Thus any values given in this paper at a specific time are actually averaged over 5~Myr.

\begin{figure}
	\centering
		\includegraphics[width=0.99\columnwidth]{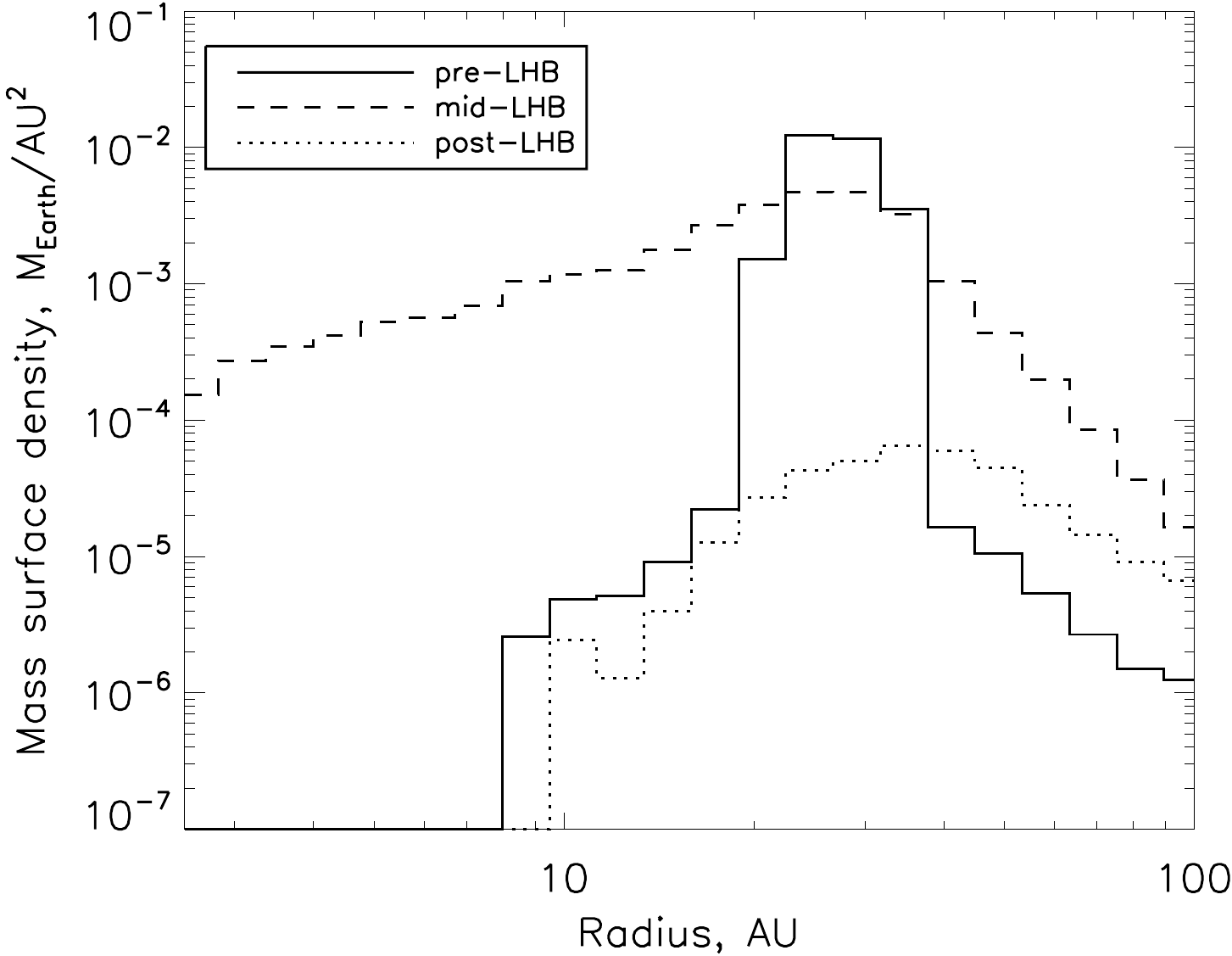}
		\caption{Mass distribution before, during and after the LHB.}
	\label{fig:lhbmvsr}
\end{figure}

Figure \ref{fig:lhbmvsr} shows the surface density of the disc: just before the LHB (at 873~Myr); during the LHB (at 881~Myr); and at the end of the simulation (at 1212~Myr). Henceforth we refer to these epochs as pre-LHB, mid-LHB and post-LHB. Before the LHB occurs, most of the planetesimals are confined to a ring $\sim$15~AU wide, centred at about 26~AU. The surface density profile interior to this ring (over the range 8-19~AU) has a slope of $R^{2.9\pm0.4}$ and outside the ring (over the range 38-106~AU) the slope is $R^{-3.2\pm0.2}$. The mass surface density within the ring is 2 orders of magnitude above this. At the onset of the LHB, a large number of the planetesimals are scattered from the belt both inwards and outwards spreading out the distribution of mass. Although planetesimals are scattered inwards before the LHB, at the onset of the LHB the rate at which planetesimals are being scattered inwards is much greater than the rate at which they are scattered back out. This results in a surface density profile with a leading slope of $R^{1.5\pm0.1}$ between 1 and 27~AU and a trailing slope of $R^{-4.8\pm0.1}$ between 27 and 106~AU at 881~Myr. By 1212~Myr the planetesimals are highly scattered giving a surface density profile with a leading slope (over the range 9-38~AU) of $R^{3.4\pm0.4}$ and a trailing slope (over the range 38-106~AU) of $R^{-2.7\pm0.1}$.

\citet{levison06} ran simulations of ecliptic comets to help them understand the orbit of the comet 2P/Encke. Their data can be used to find the spatial distribution of comets which we might expect to be similar to our distribution at the end of the Nice model. Between 9-38~AU the number density of comets is proportional to $R^{2.7}$ and between 38-106~AU the number density is proportional to $R^{-2.5}$. Our slope for the inner region is steeper than that found by \citet{levison06} and there are no particles within 9~AU whereas their comet model includes particles as far in as 0.1~AU. In part this difference is because the Nice model removes particles as soon as $q<1$~AU therefore underestimating the number of comets in this region. The slopes for the outer region compare much better despite the fact that \citet{levison06} only include objects that have interacted with Neptune in their simulations suggesting that cometary dynamics is broadly similar, at least in terms of spatial distribution.

The Nice model ends roughly 3~Gyr ago. For this work we would like to extrapolate the post-LHB evolution so that we can compare the predictions of the Nice model with current observations of the Solar System and compare our model with observations of extrasolar debris discs. To do this we need to find how the mass of the system will continue to evolve after the end of the LHB.

At the end of the Nice model run there are 322 particles remaining. Many of these particles have left the confines of the Kuiper belt, becoming comets and scattered disc objects (SDOs), with only a few remaining as classical Kuiper belt objects (CKBOs). Here we define CKBOs as objects with $q>38$ AU and 42 $<a<$ 47~AU and assume that, since these objects are now on orbits that no longer bring them close to Neptune (or any of the other planets), they would be expected to remain trapped in the classical belt for the rest of the Solar System's lifetime \citep{levison08}. In reality, the mass of the CKB will be decreased due to chaotic diffusion by resonances, small KBOs encountering large KBOs and collisions. However, this reduction in mass is a small fraction of the total mass \citep{gomes08}. From this definition we find that 3 out of the 322 particles represent CKBOs and that the mass of the classical Kuiper belt remains fixed at $M_{\rm{CKB}}=0.010\pm0.006$~M$_{\oplus}$, which compares favourably with recent observational estimates -- 0.008-0.1~M$_{\oplus}$ \citep{gladman01, bernstein04, fuentes08} -- although this may be an underestimate as the more detailed modelling of \citet{levison08}, which includes dynamical processes not present in \citet{gomes05}, gives a final mass of 0.05-0.14~M$_{\oplus}$.

\begin{figure}
	\centering
		\includegraphics[width=0.99\columnwidth]{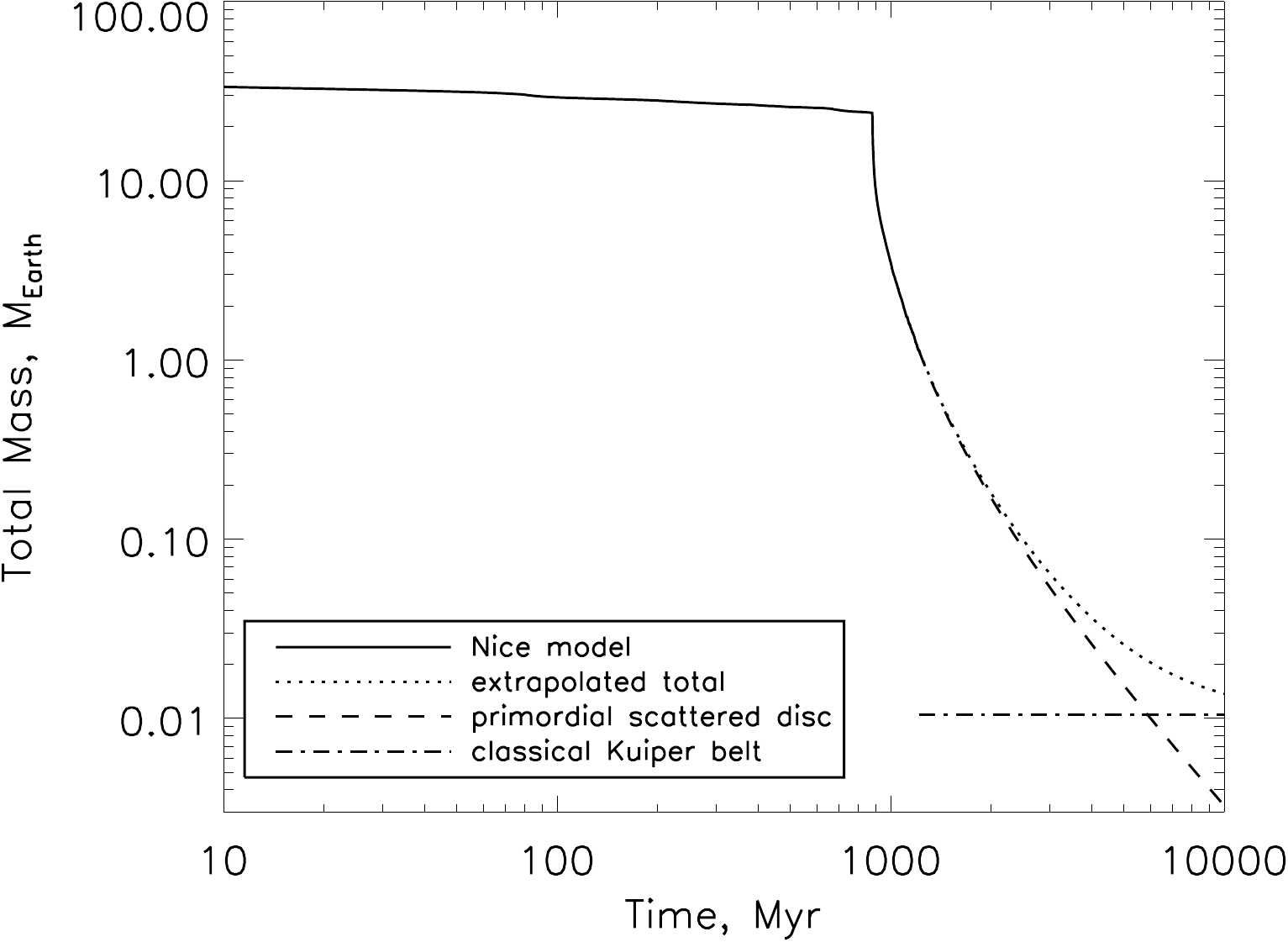}
		\caption{Total mass of Kuiper belt objects in the Nice model and extrapolated mass beyond the Nice model using equation \protect\ref{eq:mev}. The scattered disc (dashed line) and classical belt (dot-dashed line) contributions to the total extrapolated mass are also shown.}
	\label{fig:massev}
\end{figure}

As there are no major changes to the dynamical processes in the system after the LHB, we assume that the dynamical losses affecting the rest of the particles (for which we use the term primordial scattered disc) remain the same and so the total mass will continue to decline. Therefore, we can extrapolate the mass evolution to the present day and into the future (figure \ref{fig:massev}). From $\sim$950~Myr onwards, the total mass (in Earth masses) as a function of time (in Myrs) can be fitted by the equation:
\begin{equation}
M_{\rm{tot}}=\frac{3.6}{(1+(t-995)/280)^{2}}+M_{\rm{CKB}}
\label{eq:mev}
\end{equation}
which puts the current total mass at 0.03~M$_{\oplus}$ and the mass in the primordial scattered disc at 0.02~M$_{\oplus}$, which agrees well with the observations that set the current mass of the scattered disc to between 0.01-0.1~M$_{\oplus}$ \citep[and references therein]{gomes08}. The evolution of the total mass of Kuiper belt objects in the system is shown in figure \ref{fig:massev} which also shows the constant component of the CKB and the depleting component of the SD which combine to make up the total extrapolated mass.

\subsection{Converting mass to dust emission}
\label{s:m2d}
Each particle in the simulation represents a collection of Kuiper belt objects of many different sizes which are assumed to be affected by dynamical perturbations in the same way. By making some assumptions about the size distribution of the objects (which will be considered in more detail in section \ref{s:3phase}) we can work out the cross-sectional area of dust corresponding to the mass in each radius bin and thus, the flux emitted from these particles.

First we assume that a collisional cascade is set-up quickly (from $t$=0) and so the planetesimals are in collisional equilibrium with a differential size distribution of the form $n(D)\propto D^{2-3q_{\rm{d}}}$ where $D$ is the diameter of the particles (in km) and $q_{\rm{d}}=11/6$ for an infinite collisional cascade \citep{dohnanyi68}. This size distribution is assumed to apply from the largest objects of size $D_{\rm{c}}$ down to the smallest particles of size $D_{\rm{bl}}$. Particles smaller than this are created in collisions between larger objects but are blown out of the system by radiation pressure on dynamical timescales and so contribute little to the size distribution.

Taking $q_{\rm{d}}=11/6$, $D_{\rm{c}}=2000$~km, $D_{\rm{bl}}=2.2$~$\mu$m and the particle density, $\rho=1000\mbox{ kg m}^{-3}$, we find that the total cross-sectional area in each radius bin ($\sigma(R)$ in AU$^2$) is related to the total mass in each radius bin ($M(R)$ in M$_{\oplus}$) by:
\begin{equation}
\frac{\sigma(R)}{M(R)}=0.19\mbox{ AU}^{2}\mbox{ M}_{\oplus}^{-1}.
\label{eq:sigtot2}
\end{equation}

As a first approximation we assume that the particles act like perfect black-bodies, except at submillimetre wavelengths where an additional modification is applied. This is fine for large grains but does not work for small grains. Improvements to this assumption will be considered in section \ref{s:pp}. Using the black-body emission from the particles and the cross-sectional area worked out in equation \ref{eq:sigtot2} we can find the flux density (in Jy) measured by a distant observer:
\begin{equation}
F_{\nu}=\sum_R2.35\times10^{-11}\sigma(R)B_{\nu}(\lambda,T_{\rm{bb}}(R))d^{-2}X_\lambda^{-1}
\label{eq:fnu}
\end{equation}
\begin{equation}
T_{\rm{bb}}(R)=278.3 L_{\star}^{1/4}R^{-1/2}
\label{eq:tr}
\end{equation}
where $B_{\nu}$ is the Planck function (in units of Jy sr$^{-1}$), which is dependent on the wavelength and temperature, $d$ is the distance to the observer (in pc), $L_{\star}$ is the luminosity of the star (in units of L$_\odot$), $R$ is the distance between the particle and the star (in AU) and $X_\lambda$ is a factor that accounts for the drop off in the emission spectrum beyond $\sim$200~$\mu$m. Here we take $X_\lambda=1$ for $\lambda<$ 210~$\mu$m and $X_\lambda=\lambda/210$ for $\lambda\geq$ 210~$\mu$m to be consistent with submillimetre observations of extrasolar debris discs \citep[see][]{wyatt07a}. The numerical coefficient in equation \ref{eq:fnu} arises because
different units are used for different parameters, an approach employed
throughout the paper.

\begin{figure}
	\centering
		\includegraphics[width=0.99\columnwidth]{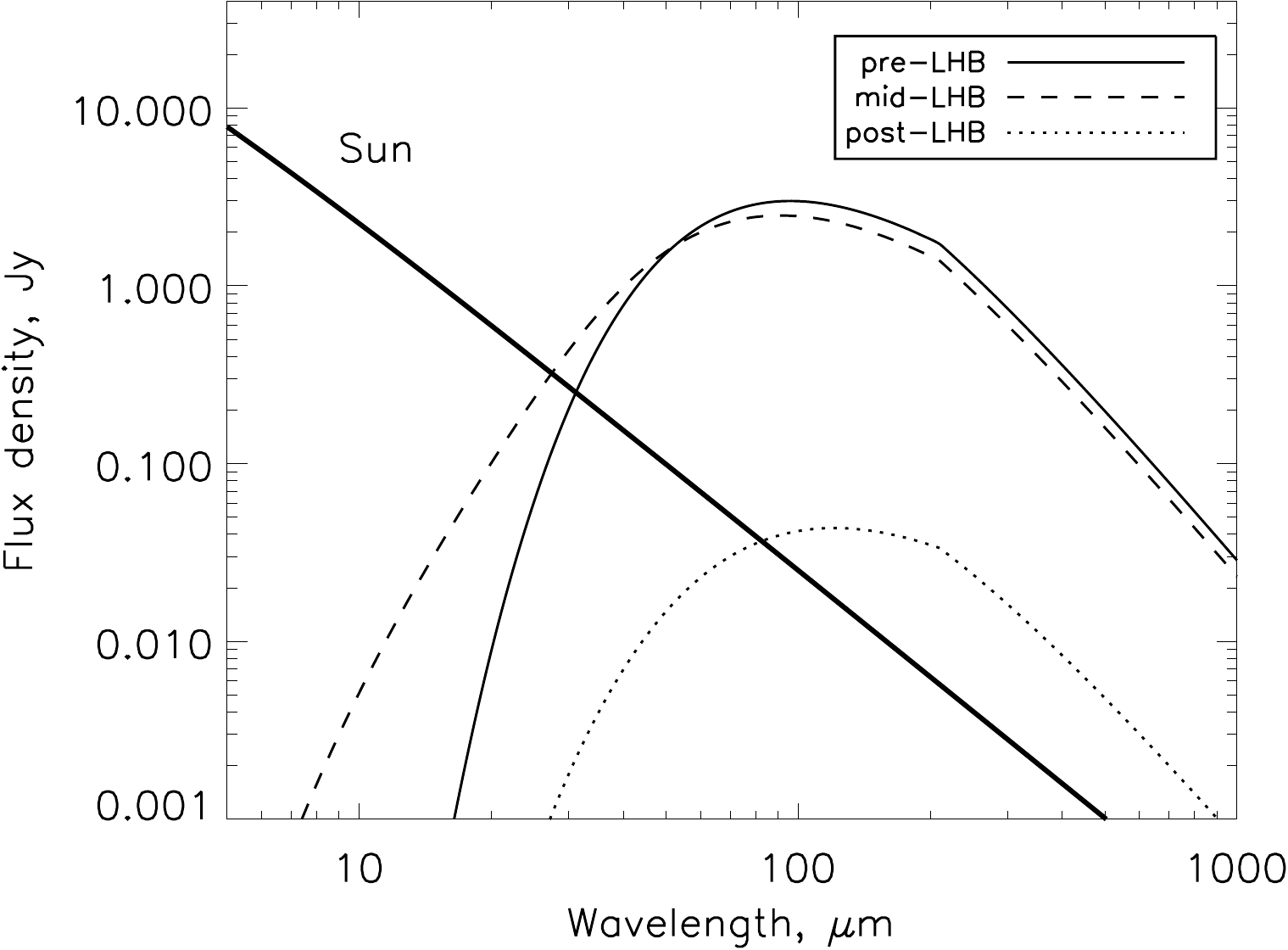}
		\caption{SED before, during and after the LHB, as it would appear from 10 pc away. The thick line shows the solar photosphere. The thin lines show the excess emission at 873 Myr, 881 Myr and 1212 Myr. The total emission spectrum observed would be the sum of the photosphere and excess.}
	\label{fig:lhbfnuvswav}
\end{figure}

By plotting the flux density against wavelength for the dust emission (see figure \ref{fig:lhbfnuvswav}) we can see how the LHB causes the emission spectrum to change. Before the LHB, the emission resembles a single temperature spectrum appropriate to the radius of the belt (i.e. 55~K at 26~AU). During the LHB, the spreading of mass from the belt (see figure \ref{fig:lhbmvsr}) means that the dust is emitting from a much broader range of temperatures and so the spectrum covers a broader range of wavelengths. Wavelengths as low as 7~$\mu$m now have a flux density $> 10^{-3}$~Jy as opposed to just wavelengths 16~$\mu$m and longer in the pre-LHB phase. In particular, we see that the mid-IR flux is enhanced. After the LHB has occurred, the flux at all wavelengths rapidly decreases and the spectrum begins to resemble a single temperature black body once again but at a longer peak wavelength due to the increase in mean radius of the belt. 

The increase in mid-IR flux seen here during the LHB is only a lower limit, since objects with perihelion less than 1 AU are removed from the simulation, which removes a lot of planetesimals that are scattered onto cometary orbits (as discussed in section \ref{s:mev}). These would otherwise contribute to the emission via dust production and sublimation. We also note that the true mid-IR emission may be higher at all times as we have only considered the contribution of the Kuiper belt and have not included the asteroid belt. These effects will be investigated in more detail in future work.

Flux density is dependent on distance from the observer to the star. To be able to compare different debris discs it is necessary to use a variable independent of distance. As stellar flux density (in Jy) is also $\propto d^{-2}$:
\begin{equation}
F_{\nu\star}=1.77B_{\nu}(\lambda,T_\star)L_\star T_\star^{-4}d^{-2}
\end{equation}
the excess ratio ($F_{\nu}/F_{\nu\star}$, also called the fractional excess) is one distance independent measure, but is dependent on wavelength. A variable independent of both distance and wavelength is the fractional luminosity, $f$, which measures the ratio of the excess luminosity (due to the dust) to the luminosity of the star:
\begin{equation}
f=\frac{L_{\rm{d}}}{L_\star}=\frac{\int{F_{\nu}\rm{d}\nu}}{\int{F_{\nu\star}\rm{d}\nu}}
\end{equation}
where $L_d$ is the luminosity of the dust and $L_\star$ is the luminosity of the star.

\begin{figure}
	\centering
		\includegraphics[width=0.99\columnwidth]{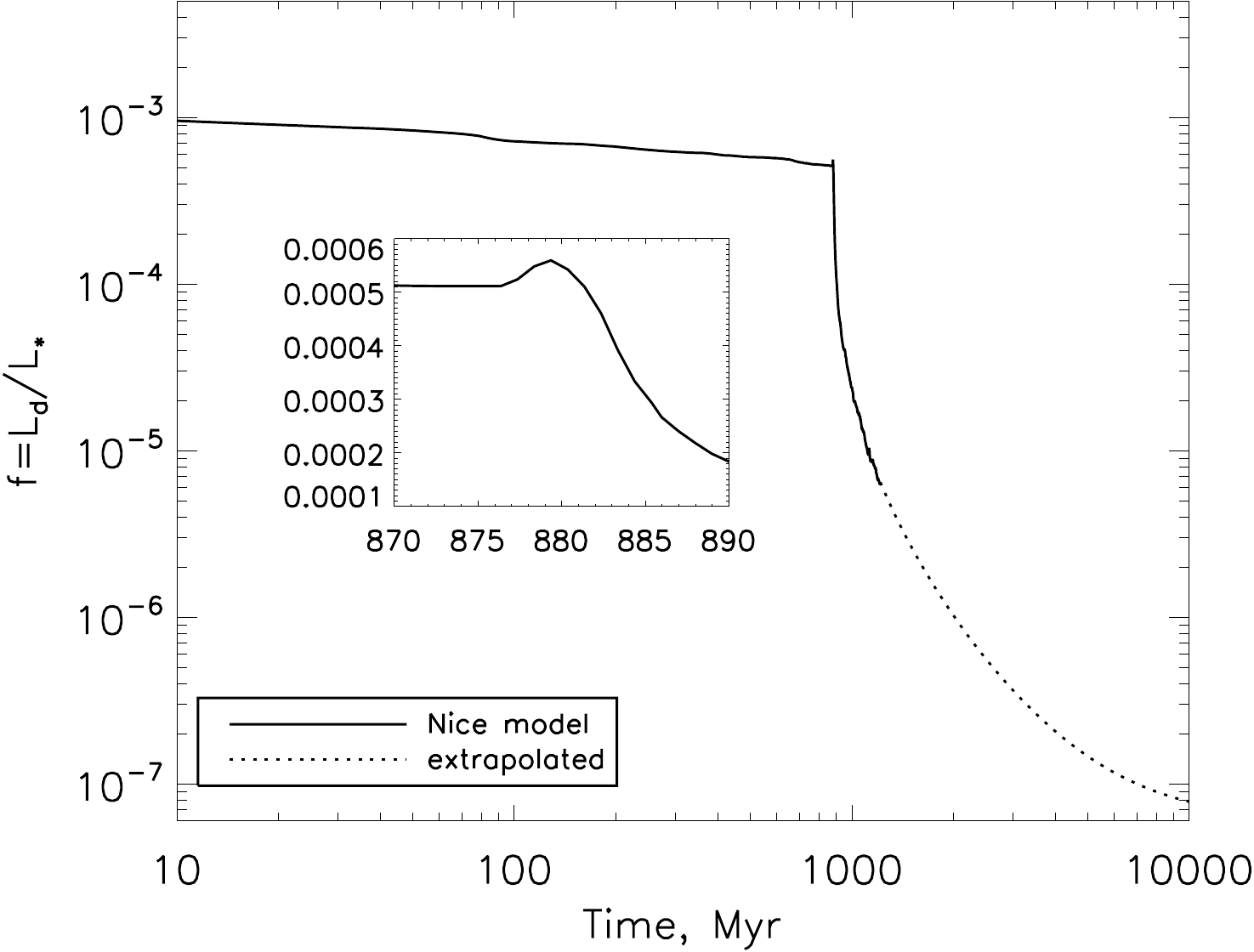}
		\caption{Fractional luminosity as a function of time. Inset shows small increase in fractional luminosity during the LHB. However, it should be noted that the peak is diminished due to the fact that we have averaged over 5~Myr (see section \protect\ref{s:mev}). The fractional luminosity at late times is an overestimate since we have ignored P-R drag and SW drag effects (see section \protect\ref{s:collpr}).}
	\label{fig:fvst}
\end{figure}

Figure \ref{fig:fvst} shows how fractional luminosity varies with time for our model. Before the LHB event, the fractional luminosity shows that the planetesimal disc is in a quasi-steady state in that dynamical losses of KBOs are a relatively small fraction of the total mass. At the time of the LHB there is a slight but minimal increase in $f$ due to the influx of comets. After the LHB, $f$ rapidly decreases in a similar manner to $M_{\rm{tot}}$ (see figure \ref{fig:massev}). Here we assume that the radial distribution of the mass remains the same from the end of the Nice model. In reality, since the mass of the CKBOs remains constant and it is only the SDOs that are being lost through dynamical processes (see section \ref{s:mev}), the distribution of mass will again resemble a narrow belt similar to that present before the LHB but at a larger radius. As such this assumption probably overestimates the mid-IR flux at late times because there would not be as much mass spread inwards as we are assuming.

By extrapolating the fractional luminosity we find that this model gives the current value to be $f=2\times10^{-7}$ which is within the range $10^{-7}-10^{-6}$ suggested by the size distribution of KBOs \citep{backman95,stern96}.

\subsection{Comparison with extrasolar debris discs}
\label{s:extradd}
Dust in debris discs emits most strongly in the infrared, as can be seen in figure \ref{fig:lhbfnuvswav}. Since its launch in 2003, the Spitzer Space Telescope has been used to survey stars for infrared excesses. The Multiband Imaging Photometer for Spitzer (MIPS) makes observations of the stars at 24~$\mu$m and 70~$\mu$m which can then be compared to photospheric models to calculate if there is evidence for any excess emission which may be due to dust present in the system. Surveys using Spitzer are generally calibration limited which means that they can detect stars with a fractional excess (the ratio of flux from the dust to flux from the star at a given wavelength) above a given limit. At 24~$\mu$m, the Formation and Evolution of Planetary Systems (FEPS) survey can make 3$\sigma$ detections of excess down to a limit of $F_{24}/F_{24\star}=0.054$ for the brightest stars \citep{carpenter09}. At 70~$\mu$m the limit is approximately $F_{70}/F_{70\star}\approx0.55$, although observations of the more distant stars are sensitivity limited and so have not be observed down to this limit \citep{wyatt08}.

\begin{figure}
	\centering
		\includegraphics[width=0.99\columnwidth]{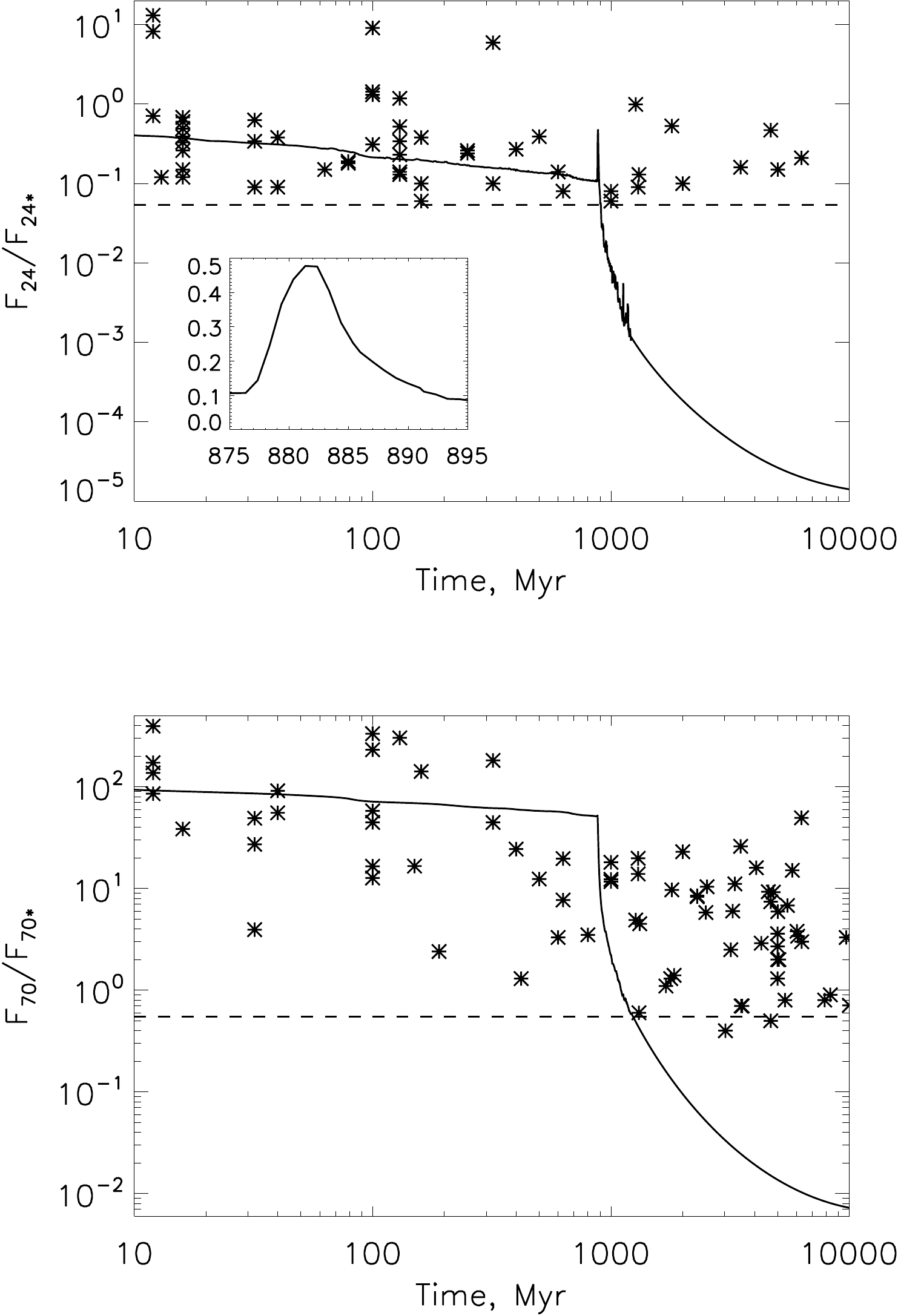}
		\caption{Excess ratio versus time for $24 \mu$m (top) and $70 \mu$m (bottom). The solid line represents the emission from our model, assuming a single-slope size distribution with $q_{\rm{d}}=11/6$ and black-body grains (c.f. figure \protect\ref{fig:f24+70rg}). The asterisks are observed discs and the dashed line shows the approximate observational limit. The excess ratio at late times is an overestimate since we have ignored P-R drag effects (see section \protect\ref{s:collpr}).}
	\label{fig:f24+70}
\end{figure}

\citet{carpenter09} surveyed 314 stars and found that there is a decrease in 24~$\mu$m excess with age. They show that 15\% of stars younger than 300~Myr have a 24~$\mu$m excess greater than 10.2\% above the photosphere but this fraction goes down to 2.7\% for older stars. By combining observations of both field stars and stars in open clusters and associations from the literature, \citet{gaspar09} also find that there is a decrease in the fraction of stars with 24~$\mu$m excess with age, levelling off at a few percent for stars older than 1~Gyr. \citet{trilling08} found that 16\% of F and G type stars have detectable debris discs at 70~$\mu$m from a sample of 225 stars. Although they show that the data could indicate a decrease in the fraction of stars with detectable excess with age, a constant excess fraction also adequately fits the data and there are currently too few observations to distinguish between the two. \citet{hillenbrand08} similarly find no apparent trend in the 70~$\mu$m excess fraction with age, however they do note that the maximum excess ratio at 70~$\mu$m does appear to decrease with age, which can be seen in figure \ref{fig:f24+70} (bottom). 

The evolution of the fractional excesses at 24~$\mu$m and 70~$\mu$m for our model are shown in figure \ref{fig:f24+70}. For comparison these plots also show 106 Sun-like stars (represented by asterisks) for which excesses have been detected \citep{habing01, beichman06a, moor06, trilling07, hillenbrand08, trilling08, carpenter09}. 77 of these stars have observed excesses at 70~$\mu$m and 53 of them have observed excesses at 24 $\mu$m. The dashed lines show the approximate limits of detectability. 

From the 24~$\mu$m excess, we can see that the hot emission from the model starts at $F_{24}/F_{24\star}=0.5$, which is high enough to make the Kuiper belt detectable at early times. This hot emission gradually decreases during the pre-LHB phase and then briefly rises again during the LHB back to its initial value (see inset of figure \ref{fig:f24+70} (top)) due to an increase in the mass closer to the Sun (see figure \ref{fig:lhbmvsr}). The Kuiper belt is also detectable at 70~$\mu$m at early times. The 70~$\mu$m excess remains in a quasi-steady state until the LHB at which point it drops off sharply, but still remains detectable up to 360~Myr after the LHB. 

It is possible that some of the observed systems may be going through a similar process and that some systems may be observed whilst in the middle of an LHB-like epoch, especially those systems described in \citet{wyatt07} as having a mid-IR excess (from dust at a few AU) higher than expected for their age. However, it is unlikely to explain systems like HD69830 which have been detected at 24~$\mu$m but not at 70~$\mu$m since our results imply that, although the 70~$\mu$m excess of a system does decrease from the time of the LHB onwards, the system should still be detectable at 70~$\mu$m for a few hundred million years after the LHB. In other words, a system must have a significant cold disc at the same time as the hot disc to provide material for the hot disc. 

Figure \ref{fig:f24+70} (bottom) shows that there are a large number of observed discs at late times, which clearly have not gone through an LHB. The fact that the \citet{trilling08} results are consistent with the
fraction of Sun-like stars with a detectable 70~$\mu$m excess remaining
approximately constant with age at $16.4^{+2.8}_{-2.9}\%$ shows that
extrasolar LHB events must be rare. Although, the number of systems
surveyed is still fairly low, we can still place an upper limit on the
fraction of systems that may undergo an LHB event. To get this limit we
start by assuming that if a star is born with a planetesimal belt that is
detectable at 70~$\mu$m then it remains detectable unless a major
planetesimal-clearing event, like a late heavy bombardment, takes place.
For the fraction of stars born with a detectable planetesimal belt we
assume that the value of $16.4^{+2.8}_{-2.9}\%$ from \citet{trilling08}
also applies at the youngest ages ($<$100~Myr). Although the Trilling
sample is not focussed on young stars, not including any systems younger
than 100Myr, the results of Carpenter et al. (2008) are consistent with
the distribution of fractional excesses remaining constant for all ages. \footnote{Note that a direct comparison of the fraction of
stars detected in each of these surveys is not possible since stars in the
different surveys were observed down to different levels of fractional
excess, notably with higher detection thresholds for the young stars in
the Carpenter survey which are typically at greater distance.} Thus the lack of
decline tells us that the fraction of stars starting with a 70~$\mu$m
excess that go through a planetesimal-clearing event is 0\% with a
3$\sigma$ upper limit of $3\sqrt{2(2.9/16.4)^2}=75\%$. This gives a maximum
of 12\% of all Sun-like stars experiencing a Late Heavy Bombardment event.

Since we might expect an LHB event to require the presence of giant
planets it is encouraging to find that this fraction is not greater than
the fraction of Sun-like stars inferred to have gas giants (planets with
masses equal to or greater than Saturn) within 20~AU which \citet{marcy05}
estimate as 12\%. If LHBs were common
for stars with giant planets then the presence of debris would be expected
to be anti-correlated with the presence of giant planets for old stars.
Since this is not observed to be the case \citep{greaves04a}
and several old stars are now known with both giant planets and debris
\citep[see table 1 in][where all the stars are at least 500~Myr]{moro07}, this is further evidence that the
fraction of stars that undergo LHBs is $<$12\% (assuming that Saturn mass
planets within 20~AU are required for an LHB).

\citet{gaspar09} also estimate the fraction of Sun-like stars that go through an LHB event. They find a maximum limit of 15-30\% based on observations of 24~$\mu$m excess. This is much higher than our own limit as they have been less restrictive with their definition of an LHB event. They assume that any star that is observed to have a 24~$\mu$m excess must be going through an LHB, however we have shown that debris discs can be detectable at 24~$\mu$m in the pre-LHB phase (see figure \ref{fig:f24+70} (top)). Furthermore, only a small number of systems have a 24~$\mu$m excess that is too high to be explained by collisional processing \citep{wyatt07}.

\begin{figure}
	\centering
		\includegraphics[width=0.99\columnwidth]{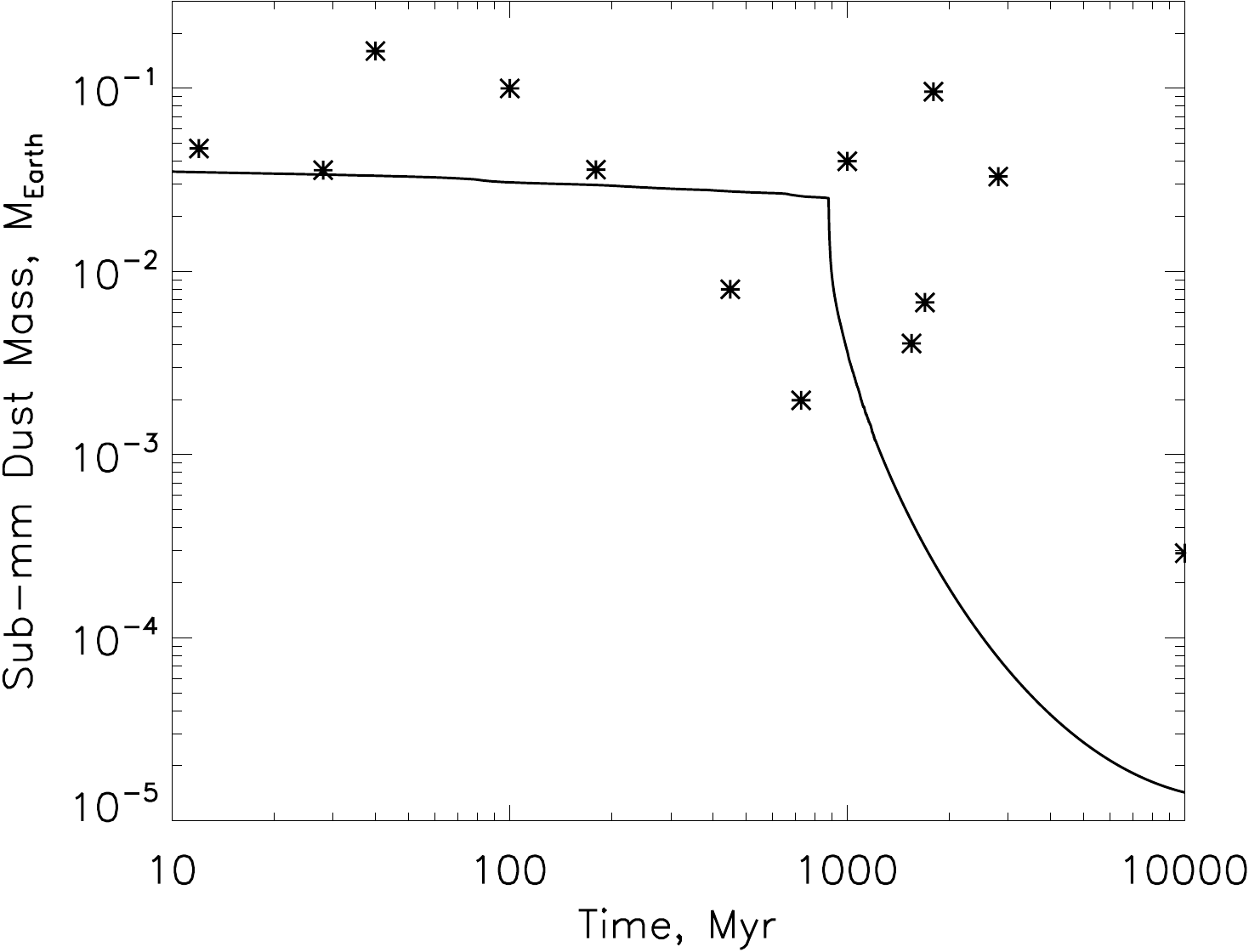}
		\caption{Sub-mm dust mass as a function of time. The solid line represents the emission from our model and the asterisks are observed discs around F, G and K stars. The dust mass at late times is an overestimate since we have ignored P-R drag effects (see section \protect\ref{s:collpr}).}
	\label{fig:submm}
\end{figure}

Observations in the submillimetre part of the spectrum offer a useful method for estimating the dust mass of debris discs and so are often used as another method of comparing debris discs. The dust mass, $M_{dust}$ in M$_{\oplus}$, can be calculated using \citep[e.g.][]{zuckerman01}:
\begin{equation}
M_{dust}=4.26\times10^{10}F_{\nu}d^2\kappa_\nu^{-1}B_\nu^{-1}
\end{equation}
where $\kappa_\nu$ is the mass absorption coefficient (in AU$^{2}$ M$_{\oplus}^{-1}$). By combining this with equation \ref{eq:fnu} we find that:
\begin{equation}
M_{dust}=\sum_R\sigma(R)\kappa_\nu^{-1}X_\lambda^{-1}.
\end{equation}
There is a lot of uncertainty in the value of $\kappa_\nu$ since it is dependant on the properties of the particles in the system. Here we adopt the value $\kappa_{850\mu m}=45$~AU$^{2}$ M$_{\oplus}^{-1}$ to ease comparison with values reported elsewhere \citep{najita05}. Figure \ref{fig:submm} shows the submillimetre dust mass predicted by our model. The asterisks represent dust masses for Sun-like stars that have been observed to have an excess at 850~$\mu$m. Data for these 13 stars is taken from the literature \citep{wyatt03a, greaves04, greaves05, sheret04, najita05, wyatt05a, williams06, greaves09}.

\citet{greaves04} use COBE/FIRAS observations at 800 $\mu$m to provide an upper limit to the dust mass of the Kuiper belt, which they find to be $\sim$2~$\times10^{-5}$~M$_\oplus$. Our model implies that the current dust mass is $\sim$3.1~$\times10^{-5}$~M$_\oplus$. The discrepancy between our result and the observations may be due to Poynting-Robertson drag being neglected in our model, which can have the effect of reducing the amount of small dust in a debris disc as described in section \ref{s:collpr}.

\subsection{Collisional lifetime}
\label{s:coll}

Mass loss in the Nice model is entirely due to the dynamical evolution of the particles. For computational reasons, it was assumed that the particles only interacted with the planets and not with each other. In reality collisional processes are likely to have had some effect on mass loss in the system. In this section we investigate the effect of collisions and P-R drag on our simple model.

In section \ref{s:m2d} we assumed that the particles are in a collisional cascade. In a collisional cascade, objects of size $D$ to $D+dD$ are destroyed by collisions only to be replaced by fragments created by collisions of larger objects. From the equations of \citet{wyatt99,wyatt07} it can be shown that the time between catastrophic collisions (known as the collision time-scale) for particles of size $D$ in a belt at a mean distance $R_{\rm{m}}$ (in AU) from the star and with a width $dr$ (in AU) is given by:
\begin{equation}
t_{\rm{c}}(D)=\left( \frac{R_{\rm{m}}^{2.5} dr}{M_\star^{0.5} \sigma_{\rm{tot}}} \right) 
            \left( \frac{2[1+1.25(e/I)^2]^{-0.5}}{f_{\rm{cc}}(D)} \right)
            \label{eq:tc}
\end{equation}
\begin{equation}
\sigma_{\rm{tot}}=\frac{\sigma(R)}{M(R)}M_{\rm{tot}}
\label{eq:sigtot3}
\end{equation}
where $t_{\rm{c}}$ is in years, $M_\star$ is the mass of the star in solar masses, $\sigma_{\rm{tot}}$ is the total surface area and $e$ and $I$ are the mean of the eccentricities and mean of the inclinations (in radians) respectively. $f_{\rm{cc}}(D)$ is a factor determined by the fraction of the total cross-sectional area which is seen by a particle of size $D$ as potentially causing a catastrophic collision and is given by:
\begin{equation}
f_{\rm{cc}}(D)=\int^{D_{\rm{c}}}_{D_{\rm{cc}}(D)}(1+D/D')^2\bar{\sigma}(D')dD'
\end{equation}
where $D_{\rm{cc}}(D)$ is the smallest particle that can catastrophically destroy a particle of size $D$ and $\bar{\sigma}$ is the normalised cross-sectional area distribution in each diameter bin. Since we are assuming a single power law with $q_{\rm{d}}>5/3$ this can be written as:
\begin{eqnarray}
f_{\rm{cc}}(D)&=&\frac{3q_{\rm{d}}-5}{D_{\rm{bl}}^{5-3q_{\rm{d}}}}\left(\frac{D_{\rm{cc}}(D)^{5-3q_{\rm{d}}}-D_{\rm{c}}^{5-3q_{\rm{d}}}}{3q_{\rm{d}}-5}\right.\nonumber \\
&&+ \frac{2D(D_{\rm{cc}}(D)^{4-3q_{\rm{d}}}-D_{\rm{c}}^{4-3q_{\rm{d}}})}{3q_{\rm{d}}-4}\nonumber \\
&&\left.+ \frac{D^2(D_{\rm{cc}}(D)^{3-3q_{\rm{d}}}-D_{\rm{c}}^{3-3q_{\rm{d}}})}{3q_{\rm{d}}-3}\right)
\end{eqnarray}
$D_{\rm{cc}}(D)=X_{\rm{c}}D$ for $X_{\rm{c}}D>D_{\rm{bl}}$ and $D_{\rm{cc}}(D)=D_{\rm{bl}}$ otherwise. The factor $X_{\rm{c}}$ can be calculated using the equation:
\begin{equation}
X_{\rm{c}}=1.3 \times 10^{-3} [Q_{\rm{D}}^\star R_{\rm{m}} M_\star^{-1} f(e,I)^{-2} ]^{1/3}
\label{eq:xc}
\end{equation}
where $Q_{\rm{D}}^\star$ is the dispersal threshold and $f(e,I)$ is the ratio of relative velocity to Keplerian velocity. This is given by:
\begin{equation}
	f(e,I) = \sqrt{1.25e^2+I^2}.
\end{equation}

Here we use the value $Q_{\rm{D}}^\star= 200$~J~kg$^{-1}$ since this value provides a good fit to the statistics of debris discs around A stars \citep{wyatt07a}. This is an effective
planetesimal strength that describes the dust mass loss rate from the
planetesimal belt which is linked to medium sized planetesimals \citep[e.g. $D_{\rm{c}}=160$~km in][]{wyatt07a}. In reality $Q_{\rm{D}}^\star$ varies with size. Thus we
expect to derive a collisional lifetime that is reasonably accurate with
regards the evolution of the infra-red emission and of the planetesimal
belt mass, but note that the collisional lifetime of objects of a specific
size will not be quantitatively correct. In section \ref{s:3phase} we investigate the effects of including a more realistic dispersal threshold that is dependent on size.

Although the planetesimals in our model are not confined to a uniform ring for the entirety of the Nice model, we can still use this model to estimate the collisional lifetime by making the assumption that $R_{\rm{m}}$ is the radius containing half of the mass of the disc, $dr$ is the annuli containing 98\% of the mass and using the mean eccentricities and mean inclinations from the Nice model data. 

The mean radius of the belt is approximately constant at 26~AU during the pre-LHB phase and then rises during the LHB due to the planetesimals being scattered and reaches 79~AU. Similarly, the width of the belt is approximately 17~AU during the pre-LHB phase and rises to a maximum of 640~AU after the LHB. The radius and width of the belt at the end of the simulation are clearly much larger than the present day classical Kuiper belt. This is because most of the objects left at the end of the simulation are scattered disc objects with a range of (typically high) eccentricities. Thus our assumption that the planetesimals are
confined to a uniform ring clearly does not hold after the LHB, and
moreover a range of collisional lifetimes would be expected depending on
the objects' orbits. Nevertheless, we note that the collision time-scale derived above is
within 25\% of that expected for an eccentric ring of planetesimals all
with semi-major axes at 102~AU and eccentricities of 0.56, which are the
mean values from the simulation at the end of the post-LHB phase \citep{wyatt09}. Thus we expect the post-LHB collisional lifetimes
presented here to be representative of an average member of the scattered
disc in this phase, but use this simply to note that the collisional
lifetime rapidly becomes larger than the age of the Solar System so that
there is no further collisional mass loss \citep[in agreement with][]{levison08a}. A consideration of collision rates in
populations with a range of eccentricities and semi-major axes \citep[see][]{wyatt09} would be required for a more detailed understanding of
collision lifetimes in the post-LHB population.

The mean eccentricities and inclinations give a mean relative velocity of around 360~m~s$^{-1}$ for the time before the LHB. At the time of the LHB, the large number of planetesimals being scattered leads to a rapid increase in the mean relative velocity, which rises to a maximum of 2700~m~s$^{-1}$ at 900~Myr. After this time, the mean relative velocity gradually decreases to a value of 2500~m~s$^{-1}$ at the end of the simulation as the highly eccentric and inclined particles are more likely to be scattered out of the system.

\subsubsection{Collisional lifetime of the largest objects}
\label{s:collar}
\begin{figure}
	\centering
		\includegraphics[width=0.99\columnwidth]{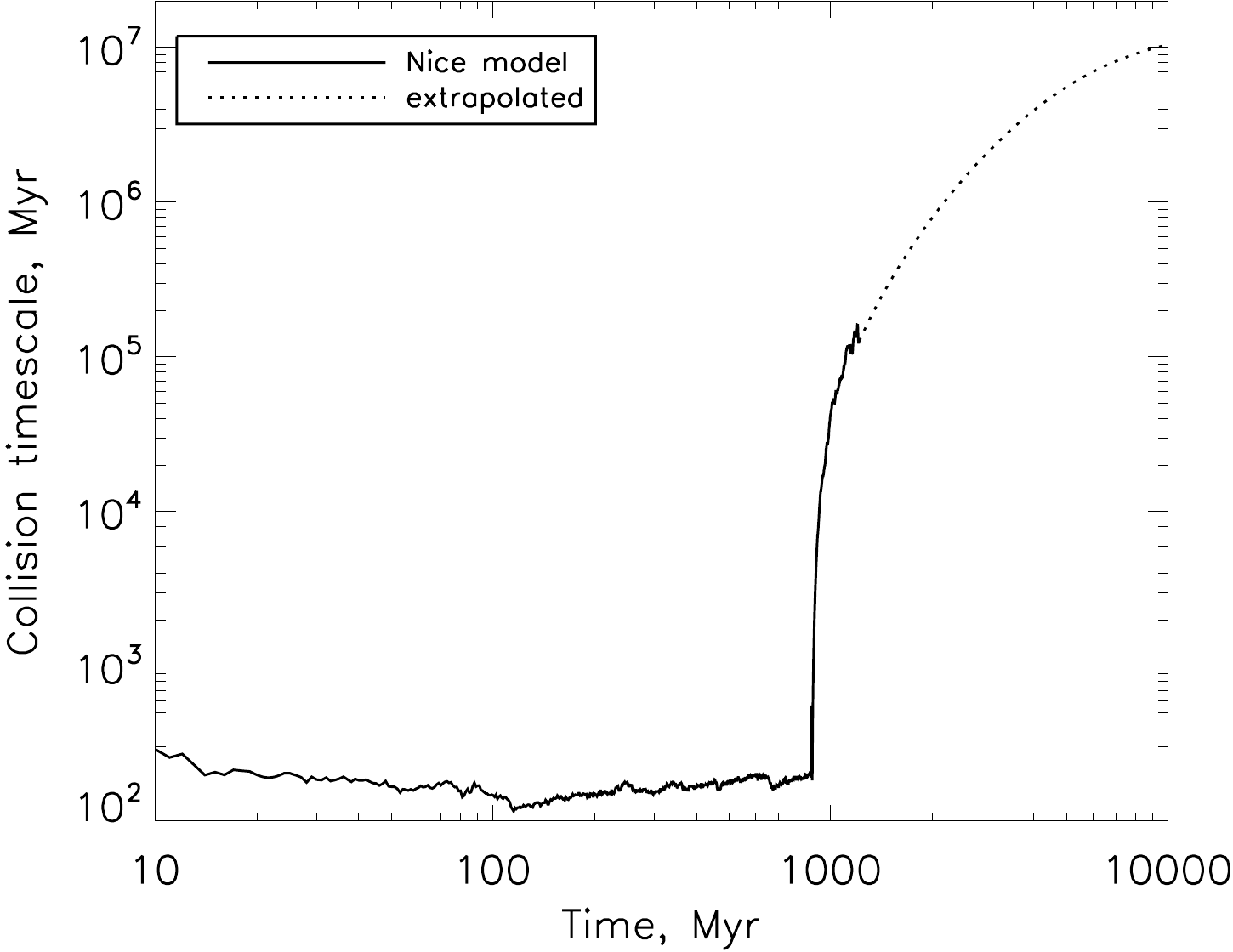}
		\caption{Collision time-scale of the largest (Pluto-sized) objects as a function of time. Catastrophic collisions are only important before and during the LHB. The actual time-scales shown here are unrealistically small due to the assumption of a size-independent dispersal threshold.}
	\label{fig:tcoll}
\end{figure}

As discussed above, $Q_{\rm{D}}^\star$ should be much greater than 200~J~kg$^{-1}$ \citep[e.g.][]{benz99} for the largest objects (of size 2000~km), however, we can still use this model to estimate qualitatively the evolution of mass due to collisions. Figure \ref{fig:tcoll} shows how the collision time-scale of the largest objects changes as the system evolves. The time-scale has been extrapolated assuming that the total mass is the only parameter in equation \ref{eq:tc} that changes with time after the end of the Nice model simulation as described in section \ref{s:mev}. During the pre-LHB period, we find that the collision time-scale varies between 100-300~Myr. This implies that catastrophic collisions might have played a significant role in mass loss before the LHB. Since we need to end up with 24~M$_\oplus$ of KBOs at the beginning of the LHB (see section \ref{s:ndata}) we can approximate how massive the initial disc must have been to account for collisional mass loss. If we assume that before the LHB all the mass was lost through collisions then the total mass evolves as \citep[e.g.][]{wyatt07}:
\begin{equation}
	M_{tot}=M_{init}/(1+t/t_c(D_{\rm{c}})).
	\label{eq:mcoll}
\end{equation}
This means that we require an initial mass of $\sim$150~M$_\oplus$ to account for the mass lost due to collisions. By taking into account the mass lost through dynamical processes ($\sim$10~M$_\oplus$) this gives us a rough estimate of 160~M$_\oplus$ as the initial mass of the Kuiper belt, much greater than the 35~M$_\oplus$ used in \citet{gomes05}. Although this is a very rough approximation of the collisional evolution and these results are not quantitatively correct due to the assumption of a size-independent dispersal threshold (which will be considered in more detail in section \ref{s:pp}) it does show us that collisions were important before the LHB and so would have affected the evolution of mass in the system.

After the instability, the collision lifetime increases to beyond the lifetime of the Solar System, with a collision lifetime of 130~Gyr by the end of the simulation and 5000~Gyr by the present day. This shows that collisions of the largest bodies become so infrequent that they can be neglected. 

\subsubsection{Lifetimes of the smallest particles}
\label{s:collpr}
\begin{figure}
	\centering
		\includegraphics[width=0.99\columnwidth]{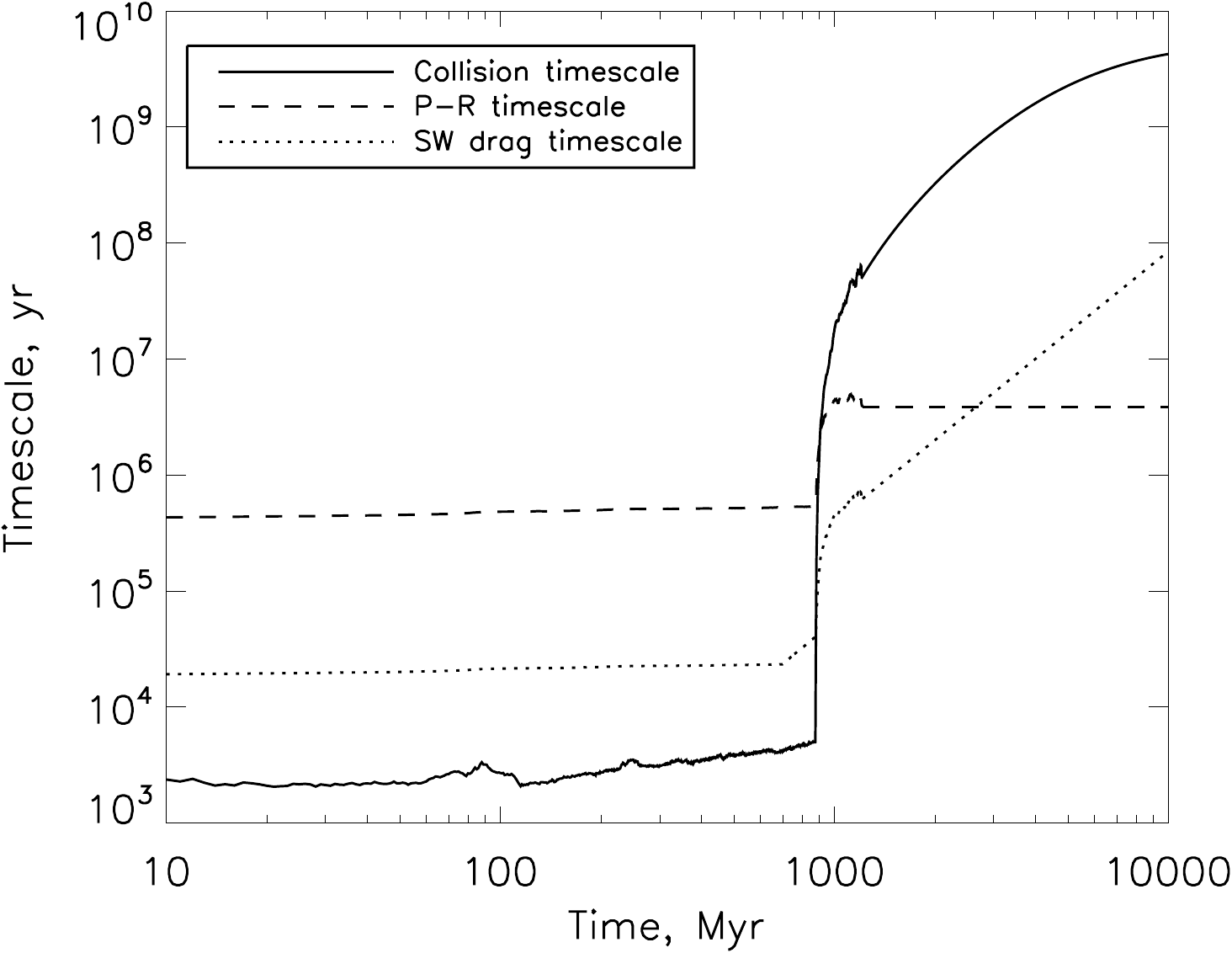}
		\caption{Collision, P-R drag and SW drag time-scales of the smallest particles (2.2 $\mu$m) as a function of time. These time-scales assume that the particles are confined to a belt with a mean radius that increases due to the scattering caused by the LHB as described in the text. The P-R drag and SW drag time-scales become important after the LHB.}
	\label{fig:tpr}
\end{figure}

In section \ref{s:m2d} we assumed that the cut-off at the small end of the size distribution was defined by radiation pressure -- particles smaller than $D_{\rm{bl}}$ will be blown out of the system by radiation pressure. However, particles larger than this may be affected by Poynting-Robertson (P-R) drag or solar wind (SW) drag on shorter time-scales than those for removal by collisions. To assess this we first calculate the collision time-scales using equation \ref{eq:tc} for particles of size $D_{\rm{bl}}$. This gives a time-scale that evolves similar to the time-scale for the largest objects but up to 5 orders of magnitude shorter (figure \ref{fig:tpr}).

P-R drag is the tangential component of the radiation force which causes a decrease in both the semi-major axis and the eccentricity of a particle. The time-scale for a particle of size $D_{\rm{bl}}$ to spiral into the Sun from a distance $R_{\rm{m}}$ under the influence of P-R drag is given by \citep{wyatt99}:
\begin{equation}
t_{\rm{P-R}}=400 \frac{M_\odot}{M_\star}\frac{R_{\rm{m}}^2}{\beta}
\end{equation}
where $\beta=0.5$ for the smallest grains.

A similar effect is caused by the tangential component of the solar wind known as the corpuscular stellar wind drag (hereafter SW drag). At the present time the SW drag force is equivalent to only 20-43\% of the P-R drag force \citep{gustafson94}, however, the higher mass loss rate of the young Sun means that SW drag would have been more effective at removing dust than P-R drag at early times \citep{minato06}. The ratio of the SW drag time-scale to the P-R drag time-scale is given by \citep{plavchan05}:
\begin{equation}
\frac{t_{\rm{SW}}}{t_{\rm{P-R}}}=3.4\frac{Q_{\rm{P-R}}}{Q_{\rm{SW}}}\frac{\dot{M}_\odot}{\dot{M}_\star}\frac{L_\star}{L_\odot}
	\label{eq:tsw}
\end{equation}
where $Q_{\rm{P-R}}/Q_{\rm{SW}}$ is the ratio of the coupling coefficients, $\dot{M}_\odot$ is the present day mass loss of the Sun and $\dot{M}_\star$ is the mass loss of the Sun at different epochs.

For this paper we have assumed that $Q_{\rm{P-R}}/Q_{\rm{SW}}=1$, $\dot{M}_\odot=2\times10^{-14}$~M$_\odot$yr$^{-1}$ and the luminosity remains constant at 1~L$_\odot$. Although the luminosity of the Sun has changed with time and is likely to have been only 0.7~L$_\odot$ when it first became a main-sequence star \citep[e.g.][and references therein]{jorgensen91}, the uncertainties in the other factors in this equation are much greater than those due to this change in luminosity. For the change in stellar mass loss rate with time we have taken \mbox{$\dot{M}_\star(t)=\dot{M}_\odot(t/4.5\:\rm{Gyr})^{-2.33}$} for $t>700$~Myr from the analysis of stellar mass loss rates in \citet{wood05} and $\dot{M}_\star(t)=$80~$\dot{M}_\odot$ for earlier times.

Figure \ref{fig:tpr} compares the collision time-scale with the P-R drag time-scale and the SW drag time-scale for particles of size $D_{\rm{bl}}$. Collisional processes would have dominated the removal of dust before the LHB due to the high dust mass present and the compactness of the belt. Drag forces would have been insignificant as they are in all other observed debris discs \citep{wyatt05}. However, as the distribution of mass becomes increasingly spread out during and after the LHB, the collision time-scales of the particles rapidly increases such that they become more susceptible to drag forces. Due to the high mass loss rate of the early Sun, SW drag is more effective at removing dust than P-R drag until $\sim$2.7~Gyr. The increased rate of dust removal due to drag forces throughout the post-LHB phase reduces the amount of small dust below that expected in the collisional cascade equation \ref{eq:sigtot3}. This means that the collisional lifetime of the smallest particles is in fact underestimated in figure \ref{fig:tpr} (which assumes the collisional cascade size distribution extends down to the blow-out limit). Since it is the smallest dust which contributes most to the emission, this increased rate of dust removal also reduces the emission and means that we have overestimated the fractional luminosity (shown in figure \ref{fig:fvst}) and the observable properties dependent on this (figures \ref{fig:f24+70} and \ref{fig:submm}) at late times. However, since it is predicted that the Kuiper belt would not be observable at this time (figures \ref{fig:f24+70} and \ref{fig:submm}), this does not affect the comparison with observed debris discs.

We note that the time-scales used here are all for the average particles in the model. In the pre-LHB phase, some particles are occasionally scattered in from the narrow belt making them more susceptible to drag forces. During and after the LHB, the range of orbital elements of the particles is greatly increased. Since both of the drag forces are proportional to $R_{\rm{m}}^2$, the position of a particle will greatly affect whether it is destroyed through collisions or spirals into the Sun due to drag forces.

We also note that the mass loss rate of the Sun at early times is not very well constrained. Some authors \citep[e.g.][]{sackmann03} have suggested that the mass loss rate of the Sun in the early Solar System may have been as much as 1000 times greater than the current value. If the mass loss rate was this high then the SW drag time-scale would have been shorter than the collisional time-scale for the smallest particles reducing the small size end of the size distribution, thus reducing the luminosity of the disc in the
pre-LHB phase below that presented here.

\section{Realistic size distribution and grain properties}
\label{s:pp}
In section \ref{s:model} we described a basic method for investigating the history of the Solar System's debris disc. That modelling method can readily be applied to the outcome of any numerical simulation to consider its observable properties. In this section we will confront two of the main assumptions of the model. So far we have been assuming that the planetesimals and dust are governed by a single phase size distribution and that they are black bodies (with a slight correction at sub-mm wavelengths).

\subsection{Three-phase size distribution}
\label{s:3phase}
\begin{figure}
	\centering
		\includegraphics[width=0.99\columnwidth]{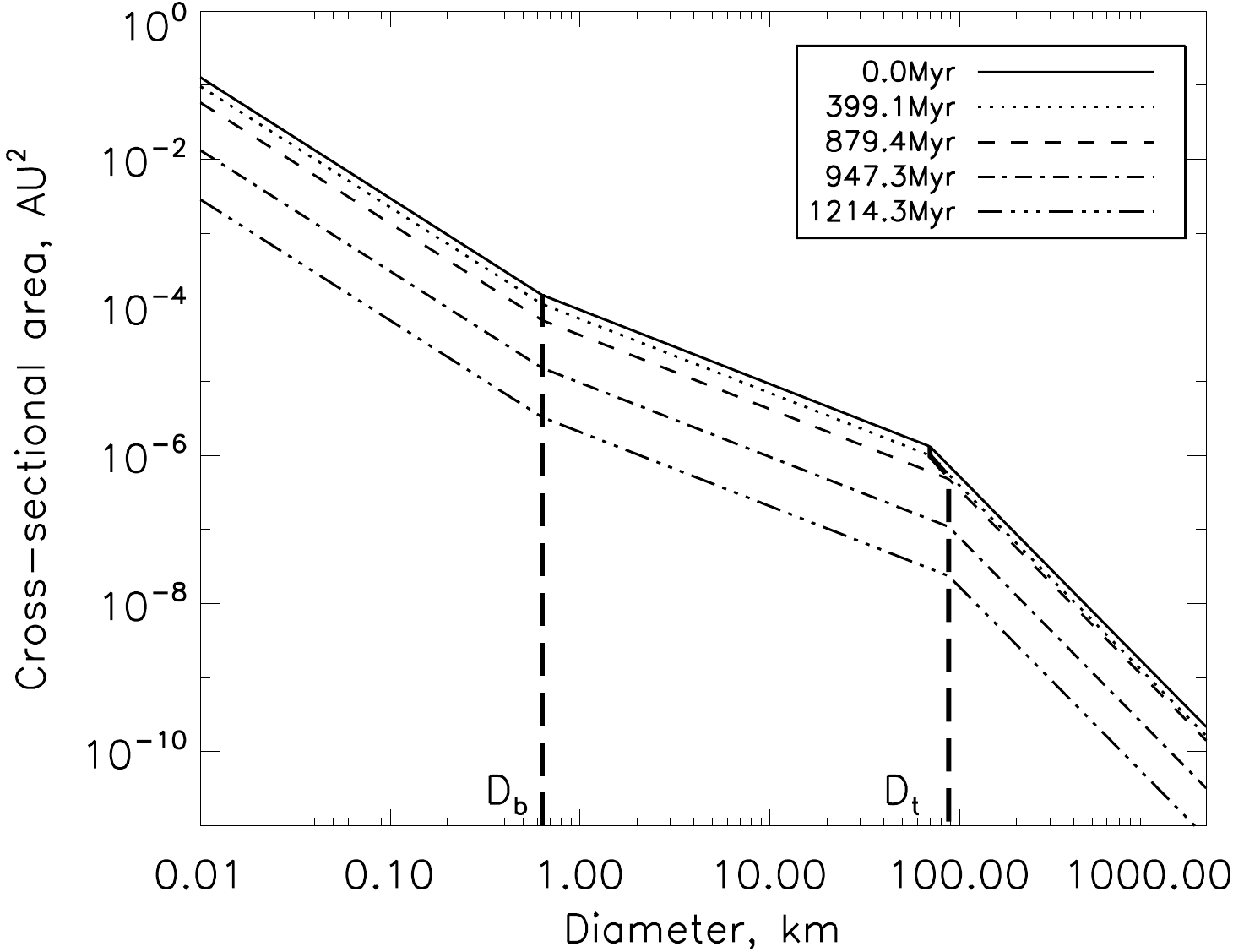}
		\caption{Evolution of the size distribution for an initial transition diameter of 70~km and initial mass of 40~M$_\oplus$. The size distribution continues down to $D_{\rm{bl}}$ but this has not been shown for clarity. The thick dashed lines represent the break diameter and the transition diameter. There is some increase in $D_{\rm{t}}$ pre-LHB when mass is being lost through collisions but the size distribution then remains fixed from the onset of the LHB and all the mass lost after this time is due to dynamics.}
	\label{fig:sigma}
\end{figure}

In section \ref{s:m2d} we assumed that the particles are in a collisional equilibrium from the smallest to the largest particle. Using their collisional evolution code, \citet{lohne08} show that a disc that starts with a single power law size distribution will quickly develop into a system with a three-phase power law due to differences in the collisional time-scales of different sized particles. As the system evolves (figure \ref{fig:sigma}), the particles begin to reach collisional equilibrium starting with the smallest particles due to their shorter collisional lifetime (see section \ref{s:coll}). The transition diameter, $D_{\rm{t}}$, defines the diameter at which the collisional lifetime of the particles is equal to the age of the system. Particles below this size will reach collisional equilibrium. The slope of the power law for particles smaller than $D_{\rm{t}}$ will then depend on whether the particles are in the strength or gravity regimes. The slope of the power law for particles larger than $D_{\rm{t}}$ is given by the primordial slope, $q_{\rm{p}}$. 

In section \ref{s:coll} we assumed that the dispersal threshold of a particle is independent of its size. This dispersal threshold defines the minimum energy required to catastrophically destroy a planetesimal and disperse the fragments such that they do not recombine under their own gravity and is, in fact, dependent on the size of the particle. The dispersal threshold decreases with size for the smallest size particles for which little energy is required to disperse the fragments after the collision \citep[the strength regime, see e.g.][]{farinella82,housen90,benz99}. As object size increases so does the gravitational strength (assuming constant density) to the extent where the energy required to disperse the fragments of a collision is greater than the energy required to catastrophically destroy the planetesimal. For these objects, the dispersal threshold increases as size increases \citep[the gravity regime, see e.g.][]{petit93,campo94,benz99}. These two regimes can be described by the sum of two power laws \citep[e.g.][]{krivov05}:
\begin{equation}
  Q_{\rm{D}}^\star(D) = A_{\rm{s}} \left(\frac{D}{2\:\mathrm{m}}\right)^{3b_{\rm{s}}} +
                A_{\rm{g}} \left(\frac{D}{2\:\mathrm{km}}\right)^{3b_{\rm{g}}}
           \label{eq:qd}
\end{equation}
where $A_{\rm{s}}$, $A_{\rm{g}}$, $b_{\rm{s}}$ and $b_{\rm{g}}$ are free parameters. This equation can then be used to find the transition between the strength and gravity regimes (the diameter at which the two power law components contribute equally), which occurs at the breaking diameter, $D_{\rm{b}}$ (in km):
\begin{equation}
D_{\rm{b}}=\left(\frac{A_{\rm{s}}}{A_{\rm{g}}}\frac{2^{3b_{\rm{g}}}}{(2\times10^{-3})^{3b_{\rm{s}}}}\right)^{1/(3b_{\rm{g}}-3b_{\rm{s}})}.
\end{equation}

To find $D_{\rm{t}}$ we assume that the disc starts as a single power law with a primordial distribution given by $q_{\rm{p}}$. Using equation \ref{eq:tc} (and adjusting $\sigma_{\rm{tot}}$ and $f_{\rm{cc}}$ for a three-phase distribution) we can find the diameter for which an object's collisional time-scale is the same as the age of the system, $t_{\rm{c}}(D_{\rm{t}})=t$. Objects smaller than $D_{\rm{t}}$ will be governed by either the strength regime (if their size is also below  $D_{\rm{b}}$) with a slope of $q_{\rm{s}}$ or the gravity regime with a slope of $q_{\rm{g}}$.

If we know the transition diameter at a particular time then we can work out the total mass of the disc at that time:
\begin{equation}
M_{\rm{totl}}(t)=8.7\times10^{-17}n_{\rm{max}}(t)\rho B(t)
\end{equation}
\begin{equation}
n_{\rm{max}}(t)=\frac{n_{\rm{max}}(0)}{1+t/t_c(D_{\rm{c}})}
\end{equation}
\begin{eqnarray}
B(t)&=&\left(\frac{D_{\rm{t}}(t)^{3q_{\rm{g}}-3q_{\rm{p}}}D_{\rm{b}}^{3q_{\rm{s}}-3q_{\rm{g}}}}{6-3q_{\rm{s}}}(D_{\rm{b}}^{6-3q_{\rm{s}}}-D_{\rm{bl}}^{6-3q_{\rm{s}}}) + \right.\nonumber \\
&&\frac{D_{\rm{t}}(t)^{3q_{\rm{g}}-3q_{\rm{p}}}}{6-3q_{\rm{g}}}(D_{\rm{t}}(t)^{6-3q_{\rm{g}}}-D_{\rm{b}}^{6-3q_{\rm{g}}}) + \nonumber \\
&&\left.\frac{D_{\rm{c}}^{6-3q_{\rm{p}}}-D_{\rm{t}}(t)^{6-3q_{\rm{p}}}}{6-3q_{\rm{p}}}\right)
\end{eqnarray}
where $n_{\rm{max}}$ is the number of objects of size $D_{\rm{c}}$ and assuming that none of $q_{\rm{s}}$, $q_{\rm{g}}$ and $q_{\rm{p}}$ are equal to 2. Given the parameters of our model, we find that the largest objects can only be destroyed by objects much larger than themselves and so cannot be collisionally destroyed in any of our simulations (see section \ref{s:res}) allowing us to ignore the $t/t_c(D_{\rm{c}})$ term. 

However, since $D_{\rm{t}}(t)$ is dependent on $M_{\rm{totl}}(t)$ (through equations \ref{eq:tc} and \ref{eq:sigtot3}), $D_{\rm{t}}$ is calculated at each time-step based on the total mass of the previous time-step. If we ignore the dynamical mass loss from the system, then the total mass at each time-step is given by:
\begin{equation}
M_{\rm{totl}}(t_n)=M_{\rm{totl}}(t_{n-1})\frac{B(t_n)}{B(t_{n-1})}.
\end{equation}
If we assume that the collisional evolution does not affect the rate of mass lost from dynamical evolution then the dynamical evolution of the Nice model (as described in section \ref{s:mev}) can be combined with this equation to give us a total mass, $M_{\rm{totl}}$, that evolves as:
\begin{equation}
M_{\rm{totl}}(t_n)=M_{\rm{totl}}(t_{n-1})\left(\frac{B(t_n)}{B(t_{n-1})}-\frac{\Delta{M}_{\rm{tot}}(t_n)}{M_{\rm{tot}}(t_{n-1})}\right)
\end{equation}
\begin{equation}
\Delta{M}_{\rm{tot}}(t_n)=M_{\rm{tot}}(t_n)-M_{\rm{tot}}(t_{n-1}).
\end{equation}

By keeping our assumption that the cross-sectional area is proportional to the mass and that the proportionality constant is defined by the size distribution, the changing size distribution (figure \ref{fig:sigma}) can then be used to find how the ratio of cross-sectional area to mass changes with time. This cross-sectional area can then be used to find the emitted flux as described for the basic model in section \ref{s:m2d}.

\subsubsection{Parameter choices}
\label{s:pchoice}
The free parameters in this model are $Q_{\rm{D}}^\star$, $q_{\rm{p}}$, $D_{\rm{t}}(0)$ and $M_{\rm{totl}}(0)$.
\citet{lohne08} use a $Q_{\rm{D}}^\star$ similar to that of \citet{benz99}. For the coefficients they set $A_{\rm{s}}=A_{\rm{g}}=500$~J~kg$^{-1}$ and the exponents are $3b_{\rm{s}}=-0.3$ and $3b_{\rm{g}}=1.5$. The transition between the strength and gravity regimes (the diameter at which the two power law components contribute equally) occurs at the breaking diameter, $D_{\rm{b}}=632$~m.

\citet{leinhardt09} show that, if comets have negligible strength, then their $Q_{\rm{D}}^\star$ function can be much lower than those of \citet{lohne08} and \citet{benz99}, with coefficients as low as $A_{\rm{s}}=20$~J~kg$^{-1}$ and $A_{\rm{g}}=28$~J~kg$^{-1}$ and exponents of $3b_{\rm{s}}=-0.4$ and $3b_{\rm{g}}=1.3$, which gives a breaking diameter of $D_{\rm{b}}=324$~m. 

\citet{stewart09} go on to show that using $Q_{\rm{D}}^\star$ is only valid when the target object is much bigger than the object impacting it. When the impacting object is roughly half the size of the target object or larger (assuming impact and target have equal density), $Q_{\rm{D}}^\star$ is no longer valid as it only takes into account the size of the target object rather than the size of both objects. In our simulations, this means that objects of size $\gtrsim$130~km become harder to catastrophically destroy and objects $\gtrsim$440~km become impossible to destroy. However, as the objects most strongly affected by this are bigger than the final transition diameter we are interested in, this does not make a significant difference to our work.

As the slope of the primordial distribution remains constant with time, this parameter can be found from present day observations. Recent surveys of the largest KBOs give a size distribution slope $q_{\rm{p}}=1.8-2.3$ \citep[see][and references therein]{petit08}. Although \citet{bernstein04} show that the classical belt and the excited belt have different size distributions, for this work we shall consider both populations to have the same size distribution. The most recent observations suggest a slope of $q_{\rm{p}}\approx2.2$ \citep{fuentes09,fraser09} and so we will use this value in the rest of this paper. $q_{\rm{s}}$ and $q_{\rm{g}}$ can be found from the formula \citep{obrien03}:
\begin{equation}
q=\frac{22/6 + b}{2+b}
\end{equation}
which sets $q_{\rm{s}}=1.877$ and $q_{\rm{g}}=5/3$.

The initial conditions, $D_{\rm{t}}(0)$ and $M_{\rm{totl}}(0)$, are less well constrained. \citet{lohne08} start their model with a single size distribution and see how it evolves in to a three-phase size distribution. As the smallest particles have such a short collisional lifetime, the size distribution is likely to have already undergone some evolution before the time at which the Nice model begins and so $D_{\rm{t}}(0)\gg D_{\rm{bl}}$. Numerical simulations of the accretion of KBOs suggests that the initial transition diameter should be no less than 1~m and probably more than 100~m \citep{kenyon99}. For the asteroid belt it has been shown that the transition diameter is likely fixed during the accretion phase \citep{bottke05}. If the same is true of the Kuiper belt then we would expect an initial transition diameter of $\sim100$~km. Models of formation of Kuiper belt objects require that there must have been at least 10~M$_\oplus$ in the primordial Kuiper belt for the largest KBOs to have formed \citep[e.g.][]{stern96a}. 

However, there are stronger constraints on the values of $D_{\rm{t}}$ and $M_{\rm{totl}}$ at the time of the LHB. The Nice model requires that there must be 24~M$_\oplus$ of planetesimals in the disc for the LHB to occur and that the size distribution becomes fixed at this time and so the transition diameter must be equal to the transition diameter in the current Kuiper belt. Observations constrain this diameter to between 50 and 200~km \citep{bernstein04,fuentes09,fraser09}. Taking these constraints into account, we can run the simulation with various initial conditions and find out which give the appropriate results. 

\begin{figure}
	\centering
		\includegraphics[width=0.99\columnwidth]{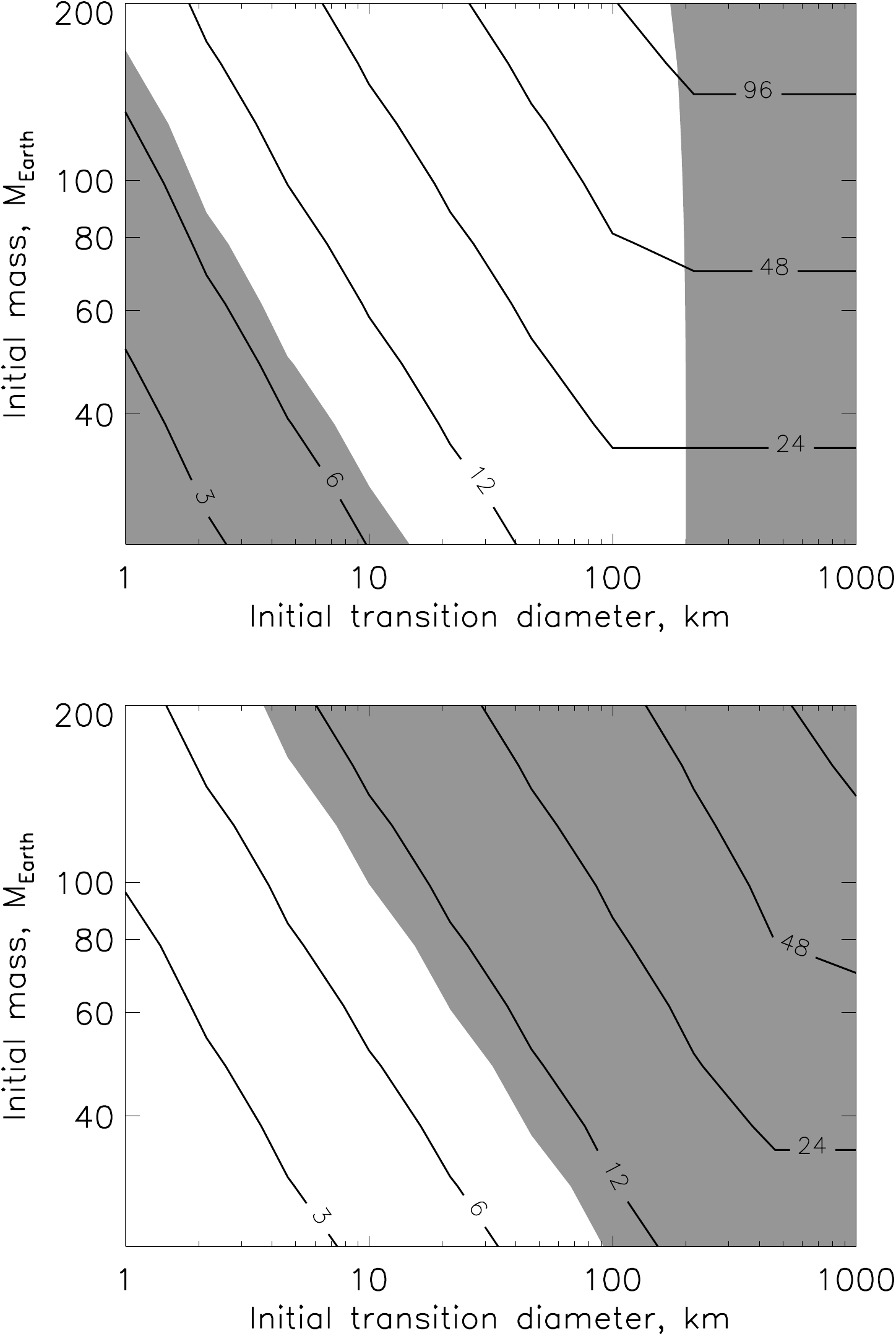}
		\caption{Dependence of mass at the time of the LHB on the initial conditions. Contours show the mass in Earth masses and the shaded regions represent runs that give a final transition diameter smaller or larger than the current constraints on the present day transition diameter. The runs shown in the top plot use the \protect\citet{lohne08} $Q_{\rm{D}}^\star$ and $q_{\rm{p}}=2.2$ and those in the bottom plot use the \protect\citet{leinhardt09} $Q_{\rm{D}}^\star$ and $q_{\rm{p}}=2.2$.}
	\label{fig:contlkr}
\end{figure}

The top plot of figure \ref{fig:contlkr} shows simulation runs using the \citet{lohne08} $Q_{\rm{D}}^\star$. The white region shows the runs which give a final transition diameter within the constraints of current observations and the 24~M$_\oplus$ line shows the runs which leave us with enough mass in the planetesimal belt for the LHB to occur. From this we can see that if we start with a transition diameter $\gtrsim$100~km then there is no collisional mass loss since objects of this size have a collision timescale longer than 879~Myr. 

The bottom plot of figure \ref{fig:contlkr} shows simulation runs using the \citet{leinhardt09} $Q_{\rm{D}}^\star$. In these runs the 24~M$_\oplus$ line does not overlap with the white area showing that we cannot satisfy both of our constraints with this $Q_{\rm{D}}^\star$. This means that the \citet{leinhardt09} $Q_{\rm{D}}^\star$ is inconsistent with the Nice model interpretation of the LHB as not enough mass remains in the Kuiper belt for the required length of time. This may be a result of our assumption that the dynamical evolution is not affected by the collisional evolution; it could mean that the objects are stronger than one would expect; or it might actually mean that the Nice model cannot be used to explain the LHB. To remain consistent with the Nice model interpretation of the LHB, we shall use the stronger $Q_{\rm{D}}^\star$ throughout the rest of this paper. We will also set $q_{\rm{p}}=2.2$, $D_{\rm{t}}(0)=70$~km and $M_{\rm{totl}}(0)=40$~M$_\oplus$ to provide an illustrative case where the mass at the time of the LHB is 24~M$_\oplus$ and the final transition diameter is 87~km.

\subsubsection{Effect of a three-phase size distribution}
\label{s:res}
In section \ref{s:model} we used a size distribution with a slope defined by $q_{\rm{d}}=11/6$ to describe all of the objects. This results in a distribution where most of the mass is concentrated in the largest objects and so the collisional mass loss of the system was dependent on the collisional time-scale of the largest objects. We now find that the largest objects of size 2000~km now require a dispersal energy of $Q_{\rm{D}}^\star(D_{\rm{c}})=1.6\times10^7$~J~kg$^{-1}$ rather than the 200~J~kg$^{-1}$ used in section \ref{s:model}. This now means the largest objects can only be destroyed by objects much larger than themselves and hence will never be destroyed in our simulations. Therefore, the largest KBOs in the Solar System are likely to be primordial as also found by the work of \citet{farinella96}. However, setting the slope of the primordial size distribution to 2.2 means that most of the mass is concentrated in objects with sizes $D\approx D_{\rm{t}}$. As $D_{\rm{t}}$ increases, mass is still lost through collisions as the primordial planetesimals reach collisional equilibrium. 

\begin{figure}
	\centering
		\includegraphics[width=0.99\columnwidth]{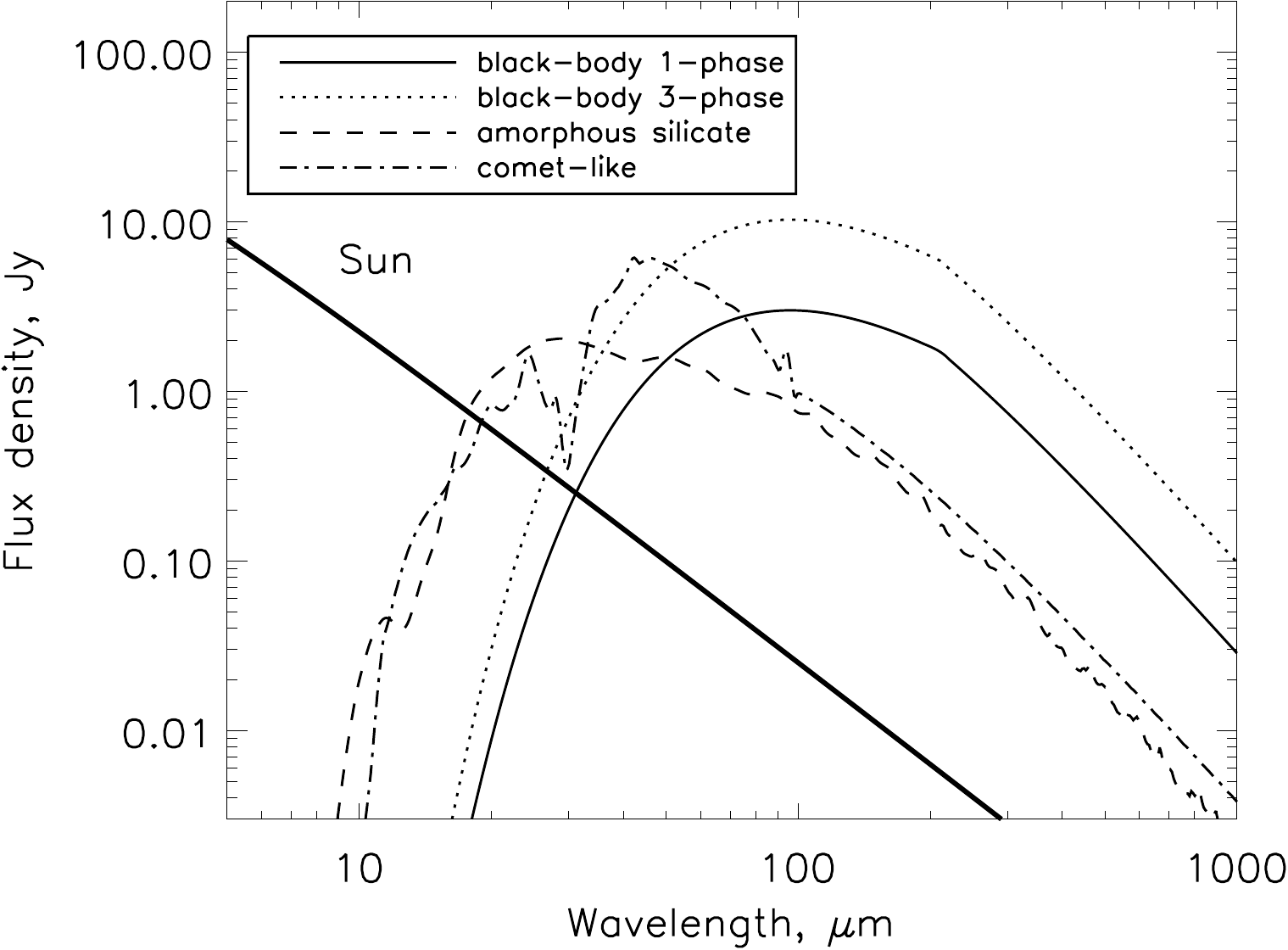}
		\caption{SED pre-LHB for black-body grains with a single slope size distribution and also black-body grains, amorphous silicate grains and comet-like grains with the L\"{o}hne model of an evolving size distribution. The realistic grain models also include a three-phase size distribution and all of the three-phase models have an initial mass of 40~M$_\oplus$ and an initial transition diameter of 70~km.}
	\label{fig:fnurg}
\end{figure}

\begin{figure*}
	\centering
		\includegraphics[width=0.99\textwidth]{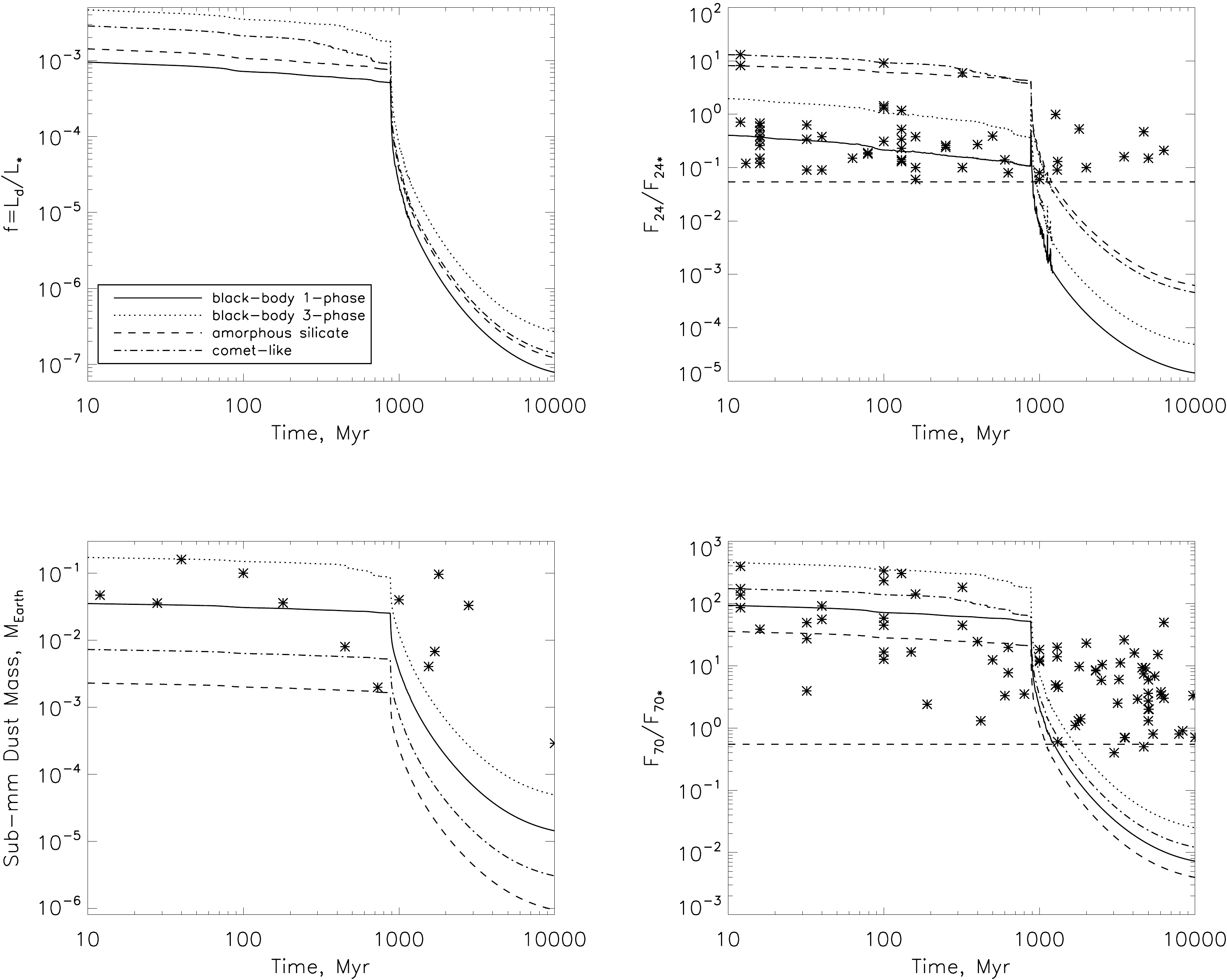}
		\caption{Same as figures \protect\ref{fig:fvst}-\protect\ref{fig:submm} but also including the black-body three-phase size distribution model and the realistic grain models. The realistic grain models also include a three-phase size distribution and all of the three-phase models have an initial mass of 40~M$_\oplus$ and an initial transition diameter of 70~km.}
	\label{fig:f24+70rg}
\end{figure*}

Figure \ref{fig:fnurg} shows how the SED changes when we use a 3-phase size distribution. By comparing the lines for black-body 1-phase and black-body 3-phase we can see that the 3-phase model increases the amount of flux emitted at all wavelengths, with a 4-fold increase in the peak of the SED. Figure \ref{fig:f24+70rg} shows how the evolution of $f$, $F_{\nu}/F_{\nu\star}$ and $M_{dust}$ changes when we use a 3-phase size distribution. From this plot we can see that the increase in flux is present at all times. This increase in flux is due to the increase in the $\sigma/M$ ratio. At early times it is five times greater than that given in equation \ref{eq:sigtot2} and decreases to four times greater as $D_{\rm{t}}$ increases. This slight decrease can be seen in the dotted lines of figure \ref{fig:f24+70rg} where the slope of the line becomes slightly steeper just prior to the LHB.

This result implies that before the LHB, the Solar System would have been amongst the brightest debris disc systems around Sun-like stars in 24~$\mu$m emission and 70~$\mu$m emission. Although plausible, this could be a result of our choice of parameters. For instance, in equation \ref{eq:qd} we set $b_{\rm{g}}=0.5$ giving us a power law slope in the gravity regime of $q_{\rm{g}}=5/3$, whereas recent observations suggest a value of $q_{\rm{g}}\approx4/3$ might be more realistic \citep{fuentes09,fraser09}.

By changing from a single slope power law to a three-phase power law and including a dispersal threshold that is dependent on size means that the Solar System's debris disc would have been significantly brighter, but the rest of the conclusions of section \ref{s:model} still hold. The 24~$\mu$m and 70~$\mu$m excesses still evolve in the same manner with a peak in the 24~$\mu$m excess still being seen at the time of the LHB. Both models also show that mass lost from collisions is likely to be important in the evolution of the Solar System's debris disc, although in the basic model the mass loss was from collisions of the largest objects whereas now we have shown that the mass loss is due to the evolution of the size distribution and the amount of mass lost is greatly dependent on the initial size distribution used.

\subsection{SED model}
\label{s:sed}
In reality, particles do not emit or absorb efficiently at all wavelengths. Their emission and absorption efficiencies are dependent on their composition and porosity. \citet{wyatt02} created a model that shows how the SED changes depending on the properties of the particles based on the compositional model of \citet{li97}. They used this model to find the most likely composition of the Fomalhaut debris disc. Here we use this model to see how it changes the results of section \ref{s:model}.

By relaxing the black-body assumption we change equations \ref{eq:fnu} and \ref{eq:tr} so that:
\begin{eqnarray}
F_{\nu}&=&2.35\times10^{-11}d^{-2}\sum_R\sum_D\sigma(R)Q_{\rm{abs}}(\lambda,D)\bar{\sigma}(D)\nonumber \\
&& \times B_{\nu}(\lambda,T(D,R))
\label{eq:fnurg}
\end{eqnarray}
\begin{eqnarray}
T(D,R)&=&(\left<Q_{\rm{abs}}\right>_{T_\star}/\left<Q_{\rm{abs}}\right>_{T(D,R)})^{1/4}T_{\rm{bb}}
\label{eq:trrg}
\end{eqnarray}
where $T_{\rm{bb}}$ is the black-body temperature given by equation \ref{eq:tr} and $Q_{\rm{abs}}$ (the absorption efficiency) is found using Mie theory, Rayleigh-Gans theory or geometric optics in the appropriate limits using optical properties from the compositional model.

To calculate the composition of the particles, the core-mantle model of \citet{li97} is used. This assumes the particles to be formed of a silicate core surrounded by an organic refractory mantle where the silicate core makes up 1/3 of the total volume. The silicate material may be either amorphous or crystalline. The particles will have a given porosity ($p$) defining how much of the particles' volume is empty space, and a given fraction of water ($q_{\rm{H_2O}}$) defining how much of the empty space is filled by ice. 

Although the composition of some of the largest KBOs has been inferred from spectroscopy \citep[e.g.][]{barucci08}, it is the smaller grains (which have not been observed) that have a larger effect on the SED. In this paper we will look at two extremes for the particle composition. The first grain composition replaces the black-body grains by amorphous silicate grains with zero porosity and no ice. For this composition $\rho=2370$ kg m$^{-3}$ and $D_{\rm{bl}}=1.47$ $\mu$m. For the second grain composition we will assume that the grains have a similar composition to the ``comet-like'' grains of \citet{augereau99}. As such, this composition uses crystalline grains with $p=0.93$ and $q_{\rm{H_2O}}=0.38$. For this composition $\rho=590$ kg m$^{-3}$ and $D_{\rm{bl}}=0.85$ $\mu$m.

\subsubsection{Effect of using a realistic grain model}
\label{s:resrg}
Figure \ref{fig:fnurg} shows how the SED for the realistic grain models compares to the black-body models from sections \ref{s:model} and \ref{s:3phase} (both realistic grain models also use the evolving size distribution described in section \ref{s:3phase}). From this plot we can see that by introducing realistic grains to the model, the peak of the SED clearly moves to a lower wavelength. This occurs because the majority of the emission is from particles in the size regime where they absorb radiation more efficiently than they emit it, which causes their temperature to increase more than if they were black-bodies (see equation \ref{eq:trrg}). The SED also shows features in the emission due to the elements present in the grain, although these are much more prominent in the ``comet-like'' model due to the presence of ice and the crystalline nature of the grains.

\begin{figure}
	\centering
		\includegraphics[width=0.99\columnwidth]{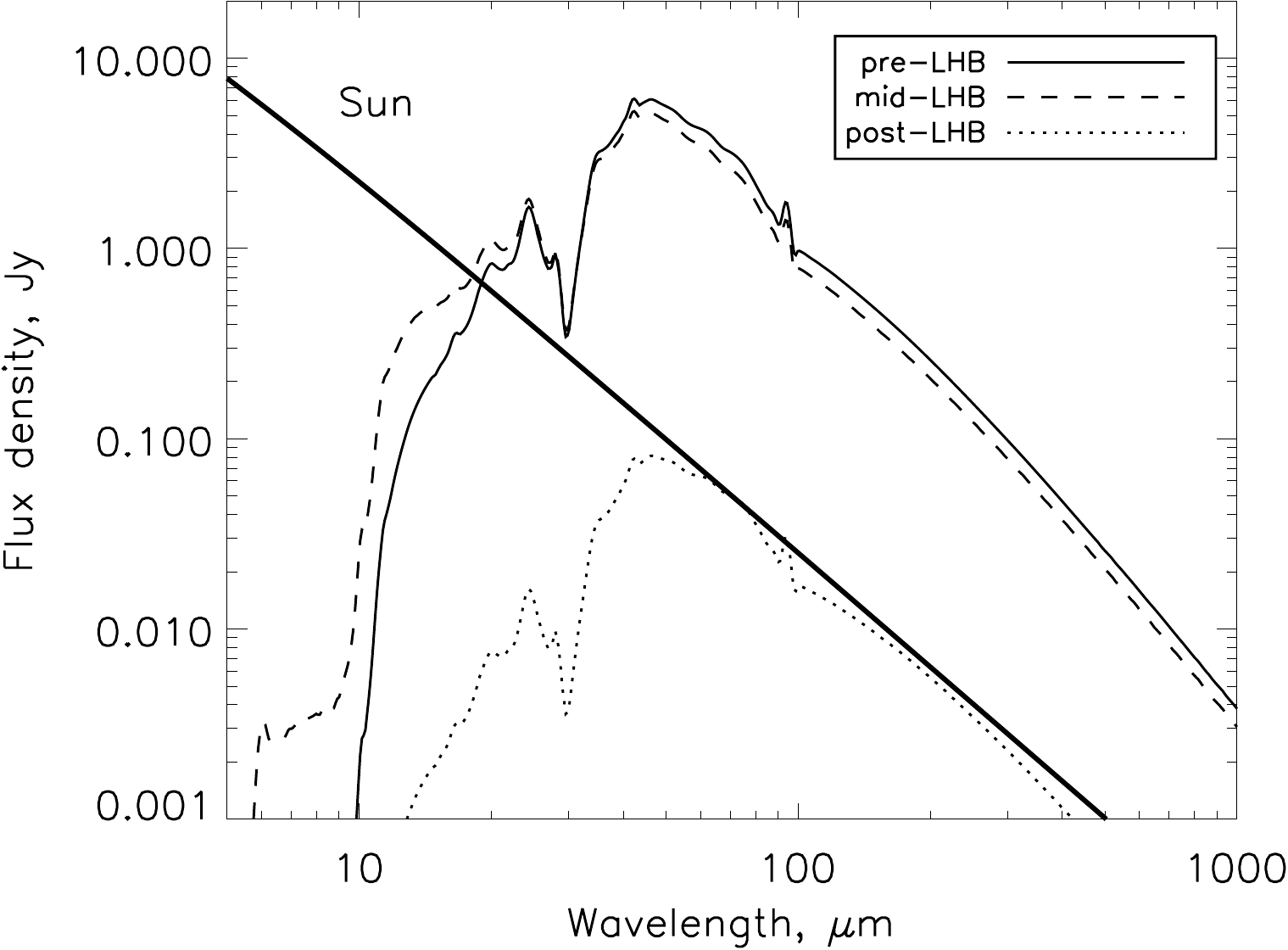}
		\caption{Same as figure \protect\ref{fig:lhbfnuvswav} but for the comet grain model.}
	\label{fig:fnuvswavc}
\end{figure}

The fact that the peak of the wavelength emission has moved to a lower wavelength means that the 24~$\mu$m flux is much higher compared to the black-body grains (see figure \ref{fig:f24+70rg}). So high that the Solar System would have been amongst the brightest discs at 24~$\mu$m before the LHB. This increase in 24~$\mu$m flux at young ages means that we no longer see the jump in $F_{24}/F_{24\star}$ that we saw for the black-body grains at the time of the LHB. However, this jump is seen at lower wavelengths, such as those below 10~$\mu$m, as can be seen in figure \ref{fig:fnuvswavc} which shows the SED for the ``comet-like'' grains before, during and after the LHB. The 70~$\mu$m flux is reduced when using realistic grain properties, although this is more pronounced for the amorphous silicate grains. This decrease is because the SED peak is at a lower wavelength and the total flux has decreased due to the decrease in emitting efficiency of the particles.

Using realistic grain models changes the density from that assumed previously, especially when porous grains are used. The lower density of the ``comet-like'' grains has the effect of increasing the $\sigma/M$ ratio which reduces the collisional lifetime of all particles. This reduction in collisional lifetime means that a greater final transition diameter is reached. For the case of the ``comet-like'' grains, a final transition diameter of 117~km is reached. However, it should be noted that we have assumed objects of all sizes to have the same density and porosity, whereas the large KBOs have lower porosities and so higher densities -- Varuna, for example, is estimated to have a porosity of 0.05-0.3 \citep{jewitt02} -- and therefore there would not be as much collisional evolution as we have suggested here. If amorphous silicate properties are used, the density of the planetesimals is much higher and so they do not collisionally evolve and the transition diameter does not change.

The high water-ice content of the ``comet-like'' model means that these particles are likely to undergo ice sublimation close to the Sun, however, the precise effects of sublimation on the size distribution are unknown. Since particles
would only come close enough to the Sun for a brief period during the LHB
(see figure \ref{fig:lhbmvsr}), we would only expect ice sublimation to affect our results
during the brief mid-LHB phase and we intend to explore this effect in a future work.

Introducing realistic grains to our model has the effect of moving the SED to shorter wavelengths, thus increasing the mid-IR flux emitted. The same effect could be approximated with black-body grains if they were assumed to be much closer to the star and therefore have a greater temperature.

Of the four models presented in this paper, the ``comet-like'' is the most realistic. This predicts that the current fractional luminosity is $3\times10^{-7}$ and the current submillimetre dust mass is $7\times10^{-6}$~M$_\oplus$. Both of which are within the observational limits (see sections \ref{s:m2d} and \ref{s:extradd}).

\section{Conclusions}
\label{s:con}
In this paper we present a new look at the history of the Solar System. Starting with the Nice model for the evolution of the Solar System, we demonstrate how the evolving spatial distribution of planetesimals in the system changes the thermal emission of dust produced in planetesimal collisions. We started with a simple model that used a single-phase power law to convert between mass of planetesimals and surface area and assumed black-body emission (see section \ref{s:model}). We find that this model predicts that the primordial Kuiper belt would have been detectable at 24 and 70~$\mu$m before the LHB. During the LHB more hot dust would have been produced as many of the KBOs are scattered inwards towards the Sun causing a peak in the 24~$\mu$m emission. Within a few hundred million years of the LHB, the dynamical depletion of the Kuiper belt renders it undetectable at 24 and 70~$\mu$m. 

Statistics from surveys of Sun-like stars \citep[e.g.][]{trilling08} show that the number of stars with an observable excess at 24~$\mu$m decreases with stellar age and the number of stars with an observable excess at 70~$\mu$m remains approximately constant with stellar age. An LHB-like event causes a drop in excess ratio of approximately 4 orders of magnitude at both these wavelengths (see figure \ref{fig:f24+70rg}) showing that such major clearing events must be rare and that most debris discs that are detectable just after 10~Myr lose mass through collisions rather than through dynamical instabilities. This allows us to set an upper limit of 12\% on the fraction of Sun-like stars that go through an LHB.

Figure \ref{fig:f24+70} shows that the period of increased 24~$\mu$m emission only lasts for $\sim$15~Myr. If we assume that the average age of Sun-like stars that we observe is 5~Gyr then there is at most a 0.04\% chance of observing a star going through a LHB. However, the bombardment from the asteroids lasts about 5 times longer \citep{gomes05} and so could increase this possibility to 0.2\%. Therefore, of the 26 Sun-like, field stars older than 10~Myr found to have a 24~$\mu$m excess out of a compiled sample of 413 \citep[see table 5 in][]{gaspar09}, approximately 1 could be an observation of a current LHB event. Certain systems such as $\eta$ Corvi which have both hot and cold dust but too much hot dust to be explained by collisional evolution alone \citep{wyatt07} may still be explained by an LHB-like event.

Although in this model we have just considered the evolution of the Solar System as described by the Nice model, this paper gives enough detail for this simple model to easily be applied to the output of any numerical simulation of planetesimals to give an indication of its observable properties.

In section \ref{s:pp} we removed the major assumptions of the initial model by including a three-phase power law size distribution that depends on collisional history. Changing the size distribution has the effect of greatly increasing the flux emitted. Having a size distribution that changes during the simulation also means that mass is also lost due to collisions. For instance, if we start with an initial transition diameter of 70~km, an initial mass of 40~M$_\oplus$ is required to leave us with the 24~M$_\oplus$ at the time of the LHB that is necessary for the LHB to take place. However, this assumes that the KBOs are as strong the \citet{benz99} case. If they are as weak as \citet{leinhardt09} suggest (and our simplistic combination of collisional and dynamical mass loss is correct) then this shows that the Nice model cannot be used to explain the LHB since either too much mass is lost through collisional grinding or the final transition diameter is unrealistic. This does not rule out the Nice model in its entirety only the need for there to be a delay before the 2:1 mean motion resonance crossing of Jupiter and Saturn.

We also find that changing the grain properties to resemble more realistic grains has the effect that the spike in 24~$\mu$m emission is no longer seen as the peak wavelength is shorter and so the 24~$\mu$m emission is initially much higher. However, peaks in emission at lower wavelengths such as 10~$\mu$m would still be possible indicators of an LHB-like, transient event. The wavelengths the spike appears at also depends on the heliocentric distance of the disc. If the disc starts further out in the system then it could still give a spike in the 24~$\mu$m emission.

One major caveat of this work is that we have only concentrated on the emission from the Kuiper Belt. Since this is far from the Sun most of the emission will be from cold dust and thus we are underestimating the warm emission and, therefore, the 24~$\mu$m flux. In future work we intend to include the asteroid belt in this model and investigate the effect of sublimation from comets.

\section*{Acknowledgements}
We would like to thank Zo\"e Leinhardt for explaining the limitations of the standard dispersal threshold model and for her comments on the paper. We also thank an anonymous reviewer for useful comments and suggestions which have helped to improve this paper. MB acknowledges the UK PPARC/STFC for a research studentship.

\bibliographystyle{mn2e}
\bibliography{thesis}{}

\bsp

\end{document}

%% file: aas_macros.tex
\def\jnl@style{\it}
\def\aaref@jnl#1{{\jnl@style#1}}

\def\aaref@jnl#1{{\jnl@style#1}}

\def\aj{\aaref@jnl{AJ}}                   
\def\araa{\aaref@jnl{ARA\&A}}             
\def\apj{\aaref@jnl{ApJ}}                 
\def\apjl{\aaref@jnl{ApJ}}                
\def\apjs{\aaref@jnl{ApJS}}               
\def\ao{\aaref@jnl{Appl.~Opt.}}           
\def\apss{\aaref@jnl{Ap\&SS}}             
\def\aap{\aaref@jnl{A\&A}}                
\def\aapr{\aaref@jnl{A\&A~Rev.}}          
\def\aaps{\aaref@jnl{A\&AS}}              
\def\azh{\aaref@jnl{AZh}}                 
\def\baas{\aaref@jnl{BAAS}}               
\def\jrasc{\aaref@jnl{JRASC}}             
\def\memras{\aaref@jnl{MmRAS}}            
\def\mnras{\aaref@jnl{MNRAS}}             
\def\pra{\aaref@jnl{Phys.~Rev.~A}}        
\def\prb{\aaref@jnl{Phys.~Rev.~B}}        
\def\prc{\aaref@jnl{Phys.~Rev.~C}}        
\def\prd{\aaref@jnl{Phys.~Rev.~D}}        
\def\pre{\aaref@jnl{Phys.~Rev.~E}}        
\def\prl{\aaref@jnl{Phys.~Rev.~Lett.}}    
\def\pasp{\aaref@jnl{PASP}}               
\def\pasj{\aaref@jnl{PASJ}}               
\def\qjras{\aaref@jnl{QJRAS}}             
\def\skytel{\aaref@jnl{S\&T}}             
\def\solphys{\aaref@jnl{Sol.~Phys.}}      
\def\sovast{\aaref@jnl{Soviet~Ast.}}      
\def\ssr{\aaref@jnl{Space~Sci.~Rev.}}     
\def\zap{\aaref@jnl{ZAp}}                 
\def\nat{\aaref@jnl{Nature}}              
\def\iaucirc{\aaref@jnl{IAU~Circ.}}       
\def\aplett{\aaref@jnl{Astrophys.~Lett.}} 
\def\apspr{\aaref@jnl{Astrophys.~Space~Phys.~Res.}}
\def\bain{\aaref@jnl{Bull.~Astron.~Inst.~Netherlands}} 
\def\fcp{\aaref@jnl{Fund.~Cosmic~Phys.}}  
\def\gca{\aaref@jnl{Geochim.~Cosmochim.~Acta}}   
\def\grl{\aaref@jnl{Geophys.~Res.~Lett.}} 
\def\jcp{\aaref@jnl{J.~Chem.~Phys.}}      
\def\jgr{\aaref@jnl{J.~Geophys.~Res.}}    
\def\jqsrt{\aaref@jnl{J.~Quant.~Spec.~Radiat.~Transf.}}
\def\memsai{\aaref@jnl{Mem.~Soc.~Astron.~Italiana}}
\def\nphysa{\aaref@jnl{Nucl.~Phys.~A}}   
\def\physrep{\aaref@jnl{Phys.~Rep.}}   
\def\physscr{\aaref@jnl{Phys.~Scr}}   
\def\planss{\aaref@jnl{Planet.~Space~Sci.}}   
\def\procspie{\aaref@jnl{Proc.~SPIE}}   